\documentclass[conference]{IEEEtran}
\ifCLASSINFOpdf
\else
\fi
\usepackage{fancyhdr}

\usepackage{algorithm}
\usepackage{xspace}
\usepackage[font=small,labelfont=bf]{caption}
\usepackage{amsmath}
\usepackage{mathtools}
\usepackage{xcolor}
\usepackage{amsfonts}
\usepackage{amssymb}
\usepackage{pifont}
\usepackage{color}

\usepackage{algpseudocode}

\usepackage{textgreek}
\usepackage{subfig}
\usepackage{textcomp}
\usepackage[]{siunitx}

\makeatletter
\algrenewcommand\ALG@beginalgorithmic{\footnotesize}
\makeatother

\usepackage{fancyhdr}
\usepackage[normalem]{ulem}
\usepackage[hyphens]{url}
\usepackage[sort,nocompress]{cite}
\usepackage[final]{microtype}
\usepackage{flushend}
\usepackage{makecell}

\usepackage[bookmarks=true,breaklinks=true,letterpaper=true,colorlinks,linkcolor=black,citecolor=blue,urlcolor=black]{hyperref}

\xspaceaddexceptions{]\}}
\newcommand{\ignore}[1]{}

\newcommand{\old}[1]{}
\newcommand{\fig}[1]{Figure~\ref{#1}}
\newcommand{\sect}[1]{Section~\ref{#1}}
\newcommand{\tab}[1]{Table~\ref{#1}}

\newcommand{\drain}[1]{\texttt{DRAIN}\xspace}

\newcommand{\nmmu}[0]{\texttt{NeuMMU}\xspace}

\newcommand{\tpr}[0]{\texttt{TPreg}\xspace}

\newcommand{\ia}[0]{\texttt{IA}\xspace}
\newcommand{\weights}[0]{\texttt{W}\xspace}

\newcommand{\prmb}[0]{\texttt{PRMB}\xspace}
\newcommand{\pts}[0]{PTS\xspace}

\renewcommand{\baselinestretch}{.999}
\title{NeuMMU: Architectural Support for Efficient Address Translations in Neural Processing Units} 

\begin{document}

\author{

\IEEEauthorblockN{
Bongjoon Hyun\hspace{2em}Youngeun Kwon\hspace{2em}Yujeong Choi\hspace{2em}John Kim\hspace{2em}Minsoo Rhu}
\IEEEauthorblockA{
School of Electrical Engineering\\
KAIST\\
\texttt{\{bongjoon.hyun, yekwon, yjchoi0606, jjk12, mrhu\}@kaist.ac.kr}\\
}
}


\maketitle
\pagestyle{plain}

\begin{abstract}
To satisfy the compute and memory demands of deep neural networks,
	 neural processing units (NPUs) are widely being utilized for accelerating
	 deep learning algorithms. Similar to how GPUs have evolved from a slave
	 device into a mainstream processor architecture, it is likely that NPUs
	 will become first-class citizens in this fast-evolving heterogeneous
	 architecture space. This paper makes a case for enabling address translation
	 in NPUs to decouple the virtual and physical memory address space.  Through
	 a careful data-driven application characterization study, we root-cause
	 several limitations of prior GPU-centric address translation schemes and
	 propose a memory management unit (MMU) that is tailored for NPUs. Compared
	 to an oracular MMU design point, our proposal incurs only an average $0.06\%$
	 performance overhead.


	\end{abstract}

\IEEEpeerreviewmaketitle


\section {Introduction}
\label{sect:intro}

The complexity of
deep neural network (DNN) based deep learning (DL) algorithms are scaling up rapidly. To meet the demands of these
computation-hungry algorithms, accelerator-centric systems based on 
GPUs or custom-designed ASICs for DNNs, often referred to as
\emph{neural processing units} (NPUs), are widely being utilized for
accelerating DL.  Similar to how GPUs have evolved into a mainstream processor
architecture, it is expected that NPUs will become first-class citizens in
heterogeneous computing platforms due to the increasing number of application
domains it is expected to accelerate.

Based on these trends, an important challenge that arises
is how NPUs should be exposed to the end-user and how the memory address space
be exposed to the NPUs.  Traditionally, the I/O attached accelerators
such as GPUs had separate memory address space than the CPU, forcing
programmers to manage these distinct address regions through manual memory
allocations, data copies, etc. As GPUs evolved into having a proper
memory management unit (MMU)~\cite{gpu_x86_at,gpu_paging,gpu_tlb}, programmers
are now given the illusion of a unified CPU-GPU memory
address~\cite{cuda_um,hsa} allowing CPU and GPU to share a globally
	addressable memory regardless of whether the physical memory is shared or
	separate. Other key features enabled by GPU MMUs include memory
	oversubscription~\cite{li:asplos:2019,nikolai:gtc:2017},
	NUMA~\cite{ararwal:asplos:2015,dgx_2,nvswitch}, and spatial sharing of a single
	GPU substrate while supporting page-granularity
	protections~\cite{ausavarungnirun:asplos:2018}. Unfortunately, these features are yet
	to be available for NPUs because they currently do not have an 
	MMU for decoupling virtual addresses against physical addresses.
	Consequently, the range of applications that
	can utilize these accelerators is limited.  For instance, NPUs 
	cannot page-fault on missing pages nor can 
	oversubscribe the NPU memory, so the working set of a
	target DNN application must precisely fit within the physical memory capacity: otherwise,
	the runtime crashes~\cite{rhu:2016:vdnn}.

\begin{figure}[t!] \centering
\includegraphics[width=0.37\textwidth]{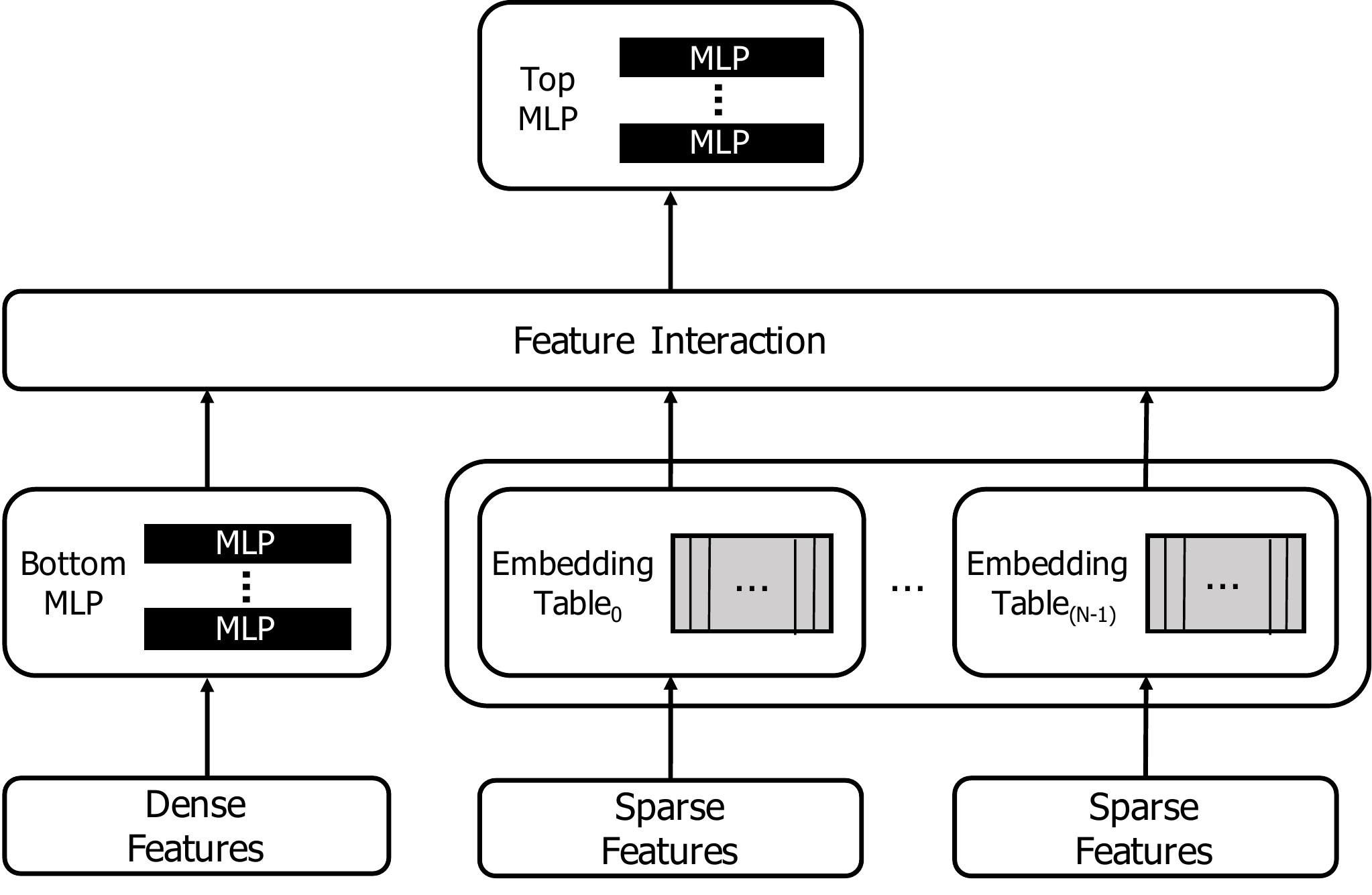}
\caption{
High-level structure of DNN-based personalized recommendation system using embeddings~\cite{facebook_dlrm,he:www:2017,mlperf}. 
Natural language processing based on attention modules (e.g., BERT~\cite{transformer,bert}), 
and memory-augmented NNs~\cite{ntm} follow a similar topological structure. That is, the application frontend starts by
\emph{gathering} multiple embeddings from large embedding lookup tables (i.e.,
		``sparse'' memory accesses),
followed by conventional, ``dense'' DNN layers (e.g., CNN, RNN, and MLP) as the backend
	processing step.
}
\label{fig:upcoming_dnns}
\end{figure}

	Given this landscape, we argue that future NPUs will need dedicated
	architectural support for \emph{virtual-to-physical address translation}
	services. Conventional wisdom in DNN's memory access characteristics is that
	they are highly regular and predictable, allowing the compiler to optimally
	decide how much memory to allocate and when/where to read/write data for NPU
	processing, obviating the need for fine-grained, page-level translations or
	NUMA capabilities. 	While such property held true for regular,
	``\emph{dense}'' DNNs (e.g, convolutional, recurrent, and multi-layer
			perceptrons, CNNs/RNNs/MLPs), emerging DL workloads employing
	\emph{embedding layers} exhibit a highly ``\emph{sparse}'', irregular, and
	random memory access pattern over a large embedding lookup table
	(\fig{fig:upcoming_dnns}).  Recent studies from Baidu and
	Facebook~\cite{hestness:2019:ppopp,park:2018:fb} state that their
	production-level DL workloads using embeddings already reached close to $100$
	GBs of memory footprint because of these embedding tables, even for inference (\sect{sect:embedding}).
	Because current NPUs incorporate \emph{only} tens of GBs of local memory,
	these memory-limited DL applications must partition its memory usage across
	CPU$-$NPU (or NPU$_{m}$$-$NPU$_{n}$ under multi-NPU
			systems~\cite{zion,habana:gaudi}), incurring frequent CPU$\leftrightarrow$NPU
	(or NPU$\leftrightarrow$NPU) data transfers.  Because of the irregular,
	random memory access nature of embedding layers, an MMU-less NPU must
	rely on the CPU to manually orchestrate data transfers on its behalf,
	experiencing significant performance overheads (\sect{sect:numa_npu}).
	Similar to how demand paging or NUMA has been a crucial component in CPUs
	(and now GPUs, especially under CPU-GPU~\cite{ararwal:asplos:2015} or
	 multi-GPU~\cite{dgx_2} systems), we believe that a robust NPU
	address translation service will enable a more diverse range of emerging,
	memory-hungry DL applications to be seamlessly executed on NPUs without
	falling into the pitfalls and performance overheads of manual management of
	NPU physical memory.

	To this end, this paper explores the design space of NPU MMUs and identifies
	and addresses the unique challenges in adding architectural support for
	address translation based on a data-driven approach. Our study consists of
	two main parts: 

	\vspace{0.5em}	
	\begin{enumerate}
	
	\item Design space exploration of an NPU MMU that enables robust address translation for conventional \emph{dense} DNN
	layers (\sect{sect:nmmu_proposal})
	
	\item Highlighting the usefulness of an NPU MMU in efficiently
	handling \emph{sparse} embedding layers via fine-grained NUMA access or page-migration (\sect{sect:numa_npu})
	
	\end{enumerate}
	\vspace{0.5em}

	As such, we start by first employing prior GPU-centric MMU
	solutions~\cite{gpu_tlb,gpu_x86_at} that utilize I/O memory management units
	(IOMMUs) to handle NPU address translations for conventional,
	dense DNNs. Interestingly, our analysis shows that, due to the
		fundamental architectural differences between GPUs and NPUs,		a naive
		IOMMU address translation incurs significant performance overhead
		even for these dense DNNs.  Concretely, while GPUs commonly use the
		on-chip SRAM for register-files and caches, NPUs almost exclusively utilize
		its SRAM for software managed \emph{scratchpads}.  The  activations and
		weights the NPUs operate on are typically multi-dimensional tensors, mapped to a 
		traditional, linear (1D) memory subsystem.  These
		tensors are much larger than the scratchpad, so the DMA unit blocks the
		activations/weights into \emph{tiles} and sequence them across multiple
		(tile) fetch operations via double-buffering.  As these tiles are also
		multi-dimensional tensors, fetching them into the scratchpad involves
		projecting the multi-dimensional coordinates into the linear space of
		DRAM memory.  A single tile is therefore decomposed into minimum number of
		\emph{linearized} memory transactions, which can be up to several
		thousands, because a tile is sized at several MBs to maximally
		utilize the scratchpad.  Consequently, a single tile fetch invokes
		significant \emph{bursts} of page translations that conventional MMUs fail
		to effectively capture, leading to an average $95\%$ performance overhead.

Overall, we observe that the bursty nature of scratchpad-based NPU address
	translation traffic renders the translation throughput of 
	baseline IOMMU's (multiple) page-table walkers (PTWs) a key performance
	bottleneck. As a result, unlike GPUs which are optimized for translation
	locality~\cite{gpu_tlb,gpu_x86_at}, we argue that \emph{NPU MMUs should be designed for
	high translation throughput first and locality second}.  To this respect, we
	propose a throughput-centric NPU MMU (\nmmu) design that effectively handles
	high burst of translation requests.  \nmmu is designed to reduce the
	address translation overhead by leveraging the deterministic memory access
	behavior of dense DNNs and the inherent translation locality therein.  Concretely,
	while TLBs are not as effective for NPUs than CPUs or GPUs, we identify how
	\emph{translation burst} locality exists within a given tile fetch in DNNs.
	To capture such intra-tile translation locality, we first propose our novel
	\emph{pending request merging buffer} (\prmb) microarchitecture as a
	translation bandwidth filtering mechanism to reduce the number of distinct
	page-table walks concurrently in-flight, boosting effective translation
	bandwidth. 	In addition to the \prmb microarchitecture, the MMU also requires
	high concurrent address translations and we evaluate the need for a larger
	number of PTWs to maximize translation throughput.  Unfortunately, the large
	amount of translations can incur significant power overheads due to the
	additional memory accesses involved in the translation process. We make the
	key observation that dense DNN layers exhibit a regular dataflow, rendering its memory access
	patterns to be highly deterministic with only a handful of key data
	structures being manipulated (i.e., input/output activations, weights). This
	allows us to employ a lightweight translation path ``register'', unlike the
	traditional MMU cache~\cite{mmu_cache}, that effectively filters down on the
	number of translation-invoked DRAM accesses (a reduction of average
			$7.1\times$). Putting everything together, while a naive IOMMU causes an average $95\%$
performance overhead,
our \nmmu effectively closes this gap,
incurring only an average $0.06\%$ performance loss (\sect{sect:nmmu_proposal}).

With our robust \nmmu design in hand, we then demonstrate the usefulness of MMU's address
	translation feature for handling embedding layers. We show that an MMU-less NPU suffers from (redundant)
		manual data copy operations between CPU and NPUs, leading to an average
		$71\%$ performance loss when executing embedding layers. Our \nmmu again
		effectively closes this performance gap using direct NUMA accesses across remote memory regions,
	achieving significant performance improvements than an MMU-less NPU design (\sect{sect:numa_npu}).
 \noindent To summarize our {\bf key contributions}:

\vspace{0.2em}
\begin{itemize}			

\item To our knowledge, this work is the first to explore architectural support
for NPU MMUs,  a significant first step in exploring this emerging design 
space.

\item We conduct a detailed, data-driven analysis on conventional (dense) DNNs and emerging
(sparse) embedding layers, root-causing the limitations of GPU-centric IOMMUs in
handling the bursty address translations of NPUs. 

\item
We propose our throughput-centric \nmmu design
based on our novel \emph{pending request merging buffer},
a high throughput parallel page-table walker, and a translation
path register, only incurring an average $0.06\%$ performance loss.

\item
Using DNN-based recommendation system models as  driving examples, we
showcase how efficiently sparse embedding layers can be handled using
fine-grained NUMA or page-migrations, enabled by \nmmu.

\end{itemize}

\section{Background \& Methodology}
\label{sect:methodology}

\subsection{Address Translation in SPM-centric NPUs}
\label{sect:npu_arch}

\begin{figure}[t!] \centering
\includegraphics[width=0.28\textwidth]{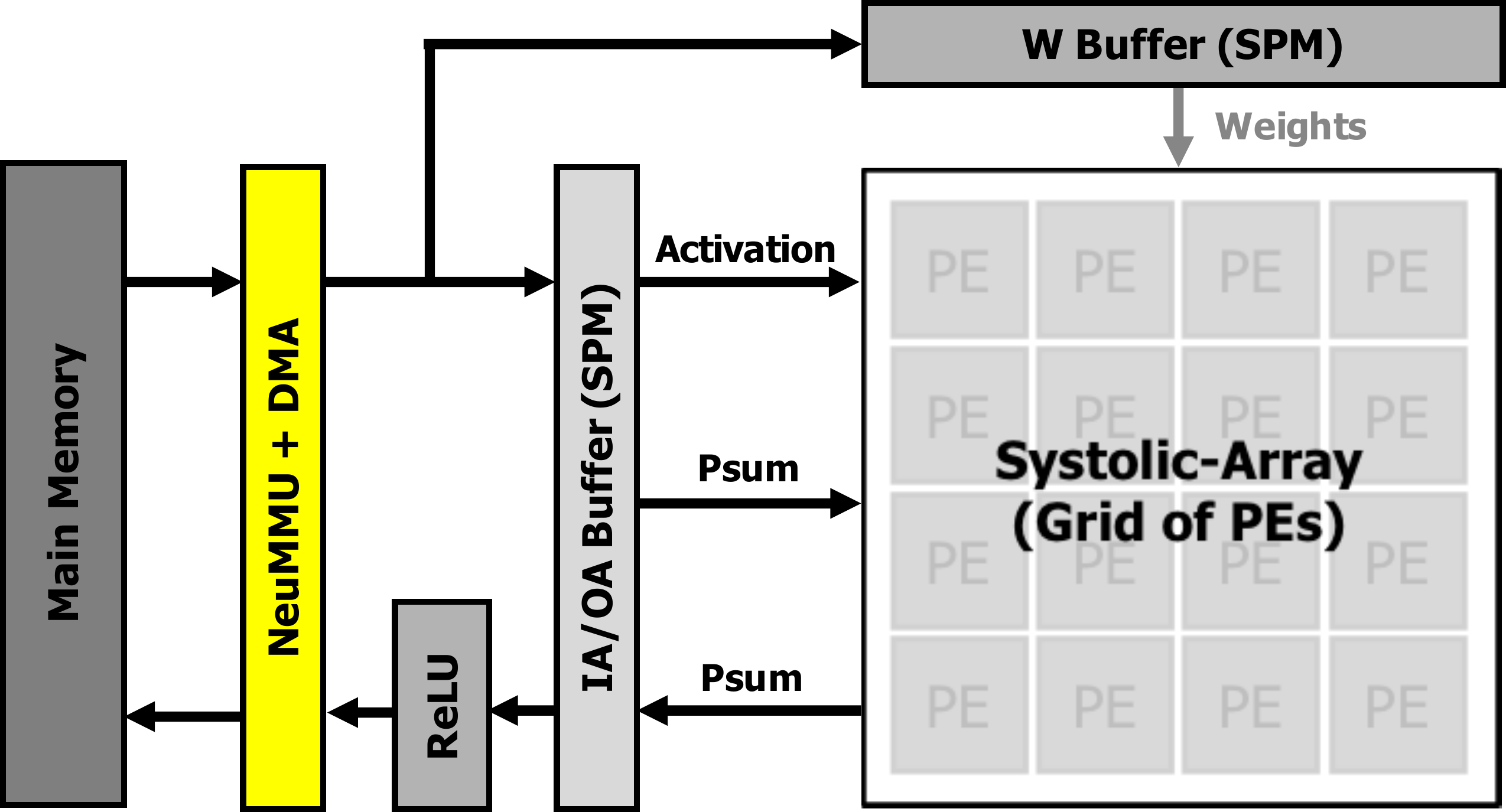}
\caption{
Baseline NPU architecture using a Google TPU-style systolic-array~\cite{tpu1}. 
	\sect{sect:eval_methodology} details the baseline NPU design.
}
\vspace{0.5em}
\label{fig:npu_dataflow}
\end{figure}

\begin{figure}[t!] \centering
\includegraphics[width=0.42\textwidth]{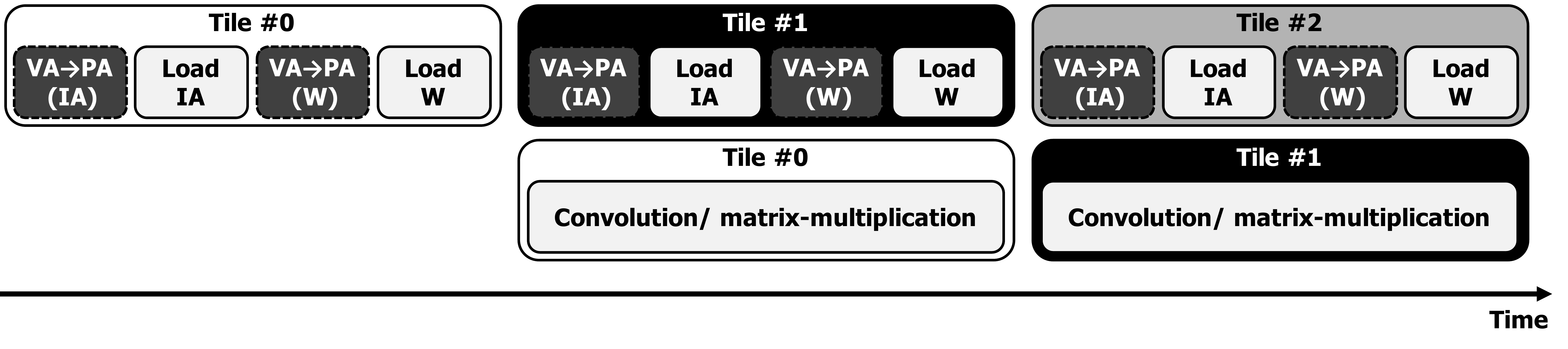}
\vspace{-0.3em}
\caption{
	Compute phase is defined
as the time NPU spends conducting the necessary computations for a given tile,
	 while memory phase is defined as the period bringing in the \ia/\weights tiles
		 into SPM.
Overlapping \texttt{tile$_{(n)}$}'s compute phase with \texttt{tile$_{(n+1)}$}'s
memory phase help maximize NPU resource utilization. 
With an NPU MMU, an address translation
of tile \ia/\weights must precede in order to fetch the actual data.
}
\label{fig:tiling_timeline}
\end{figure}

NPUs generally utilize most of its on-chip SRAM as a scratchpad memory
(SPM\footnote{While it is possible for future NPUs to adopt user-transparent
 caches, we argue that SPMs are a more natural fit for NPUs. This is
 because DNNs exhibit a highly deterministic dataflow
 (i.e., locality and data reuse information is statically available), making
 them more amenable for optimizations using the software-managed scratchpads
 for maximal resource utilization.  }) whereas GPUs allocate the bulk of this
space as register-files in order to spawn as many threads as possible for
latency tolerance~\cite{cuda} (\fig{fig:npu_dataflow}).  In contrast, NPUs commonly leverage task-level
parallelism to \emph{double-buffer} the SPM so that the latency to fetch the
input activations (\ia) and weight filters (\weights) is hidden inside the
latency to execute a layer.  Because the size of \ia and \weights can be
hundreds to thousands of MBs, the DMA unit blocks the \ia and \weights into
smaller \emph{tiles} and sequence them in and out of the SPM (e.g., typically
		tens of MBs in state-of-the-art NPUs~\cite{tpu1,volta_v100}) across
multiple iterations (\fig{fig:tiling_timeline}). A key reason why NPUs prefer a
SPM is because of the predictable performance it delivers: once the
data is brought in from memory to the SPM, the latency to read/write
data from/to the SPM is much more deterministic than caches (i.e., SPM hit rate
		is $100\%$).  From an MMU standpoint, address translation is not required
when the processing elements (PEs) access the SPM 
during the compute phase for layer computations.  In other words, the PEs
need not have to query an MMU for address translations when accessing the SPM.
During the memory phases however, the DMA unit does require information
regarding where inside the NPU physical address space the \ia and \weights are
located, as detailed below. 

\subsection{NPU Programming Model and IOMMUs}
\label{sect:npu_prog}

{\bf Current NPU programming model.} NPUs generally feature a private,
	physically-addressed memory. Consequently, the CPU must explicitly copy
	the necessary data structures (e.g., \ia and \weights) from the host memory
	to the NPU (physical) memory address space. After the CPU$\rightarrow$NPU data
	transfer is complete, the NPU-side DMA unit is given the target
	layer's \ia and \weights mapping information within the NPU physical address
	space.  Concretely, the DMA unit is given the base (\texttt{base}) and the
	boundary value (\texttt{bound}) of the data allocated, which is utilized to
	derive the physical address of target data elements, obviating the need for a
	separate address translation. Such approach is similar to the ``old''
	GPGPU programming model, which suffers from the same problems users
	had to face: 1) the working set \emph{must} fit within the NPU
	physical memory, preventing DNNs that oversubscribe NPU memory from being
	executed (e.g., large batch DNNs~\cite{rhu:2016:vdnn}), and 2) it
	becomes challenging to support ``pointer-is-a-pointer'' semantics, reducing
	programmability and complicating situations where the CPU and NPU (or
			among multiple NPUs~\cite{tpu2}) share data. To tackle these
	limitations, the I/O MMU (IOMMU)~\cite{iommu2} can be
	utilized to service accelerator-side virtual-to-physical address translations
and overcome the aforementioned limitations.
			
{\bf IOMMU hardware/software architecture.} The IOMMU is assigned with the
access privilege to walk the CPU's page-tables, allowing the CPU and the
(GPU/NPU) accelerators to share a unified global address space.  When the
accelerator is not able to locate a proper translation for its virtual address
(VA), a translation request is sent as an ATS (address translation service)
	request packet over PCIe to the IOMMU. The IOMMU can include its own TLB
	hierarchy (called IOTLB), which is checked first when an ATS is received.
	When IOTLB misses, a hardware page-table walker (PTW) inside the IOMMU walks
	the CPU page-tables to retrieve the translated physical address (PA).
	Because a single IOMMU block is designed to be shared by multiple
	accelerators (e.g., GPUs, DSPs, ISPs, and NPUs), current IOMMUs employ
	\emph{multiple} PTWs (typically $8$ but can be up to $16$), allowing multiple
	translations in-flight.

\subsection{Evaluation Methodology}
\label{sect:eval_methodology}

{\bf Baseline NPU architecture.} Our baseline NPU architecture assumes a Google
TPU-style systolic-array microarchitecture (\fig{fig:npu_dataflow}), which we
modeled as a detailed, cycle-level performance simulator by cross-referencing
publicly disclosed documents and patents from
Google~\cite{tpu1,tpu_patent1,tpu_patent2,tpu_patent3,tpu_patent4}.  Our
performance model is cross-validated against Google Cloud
TPU~\cite{cloud_tpu_beta}, achieving an
		average $80\%$ correlation in terms of effective throughput. The
		baseline NPU model employs a SPM based on-chip memory hierarchy and a
		\emph{weight-stationary dataflow}~\cite{eyeriss_isca}, as implemented in
		the original TPU (\tab{tab:npu_config}).  Similar to discrete NPUs such as Intel-Nervana's Neural
		Network Processor~\cite{nervana} or Habana's Gaudi~\cite{habana:gaudi}, our
		NPU model utilizes a local, high-bandwidth memory (e.g., HBM~\cite{hbm}).
		Similar to prior work~\cite{cnvlutin,stripes,scnn,eyeriss}, we modeled the
		memory system as having fixed latency and bandwidth rather than employing a
		cycle-level DRAM simulator~\cite{dramsim2,usimm} to reduce simulation time.
		When modeling IOMMUs, we assume an x86-64 style, hierarchical $4$-level
		page-tables with key configuration parameters following those from prior
		related literature~\cite{gpu_tlb,gpu_x86_at,cong:2017:hpca}.
		While the	remainder of this paper assumes a systolic-array based NPU for our
		discussions, the effectiveness of our \nmmu design remains intact for other
		NPU designs, such as \emph{spatial}-array based
		microarchitectures~\cite{dadiannao,eyeriss,cnvlutin,scnn,park:2018:olaware}
		as these NPUs are also based on an SPM-centric memory hierarchy. We discuss
		the implication of alternative NPU architectures and DNN dataflows	on our
		MMU proposal in \sect{sect:neummu_with_other_npus}.

\begin{table}[t!]
  \centering
	\vspace{-0.5em}
  \caption{Baseline NPU configuration.}
\scriptsize
\vspace{-0.5em}
  \begin{tabular}{|c|c|}
		\hline
		\multicolumn{2}{|c|}{\textbf{Processor architecture}} \\
		\hline
    Systolic-array dimension     			& $128 \times 128$   \\
    \hline              
    Operating frequency of PE				& $1$ GHz \\
    \hline              
    Scratchpad size (activations/weights)    			& $15$/$10$ MB \\
    \hline              

    \multicolumn{2}{|c|}{\textbf{Memory system}} \\
    \hline
    Number of memory channels 	& $8$	\\
    \hline
    Memory bandwidth	& $600$ GB/sec	\\
    \hline
    Memory access latency & $100$ cycles  \\
		\hline
    
		\multicolumn{2}{|c|}{\textbf{IOMMU}} \\
    \hline
    Number of TLB entries	& $2048$ \\
    \hline
    TLB hit latency		& $5$ cycles  \\
    \hline
    Number of page-table walkers 	& $8$	\\
    \hline
    Latency to walk page-tables  	& $100$ cycles per level	\\
		\hline

		\multicolumn{2}{|c|}{\textbf{System Interconnect}} \\
    \hline
    NUMA access latency across sytem interconnect	& $150$ cycles \\
    \hline
		CPU$\leftrightarrow$NPU Interconnect Bandwidth		& $16$ GB/sec  \\
    \hline
    NPU$\leftrightarrow$NPU Interconnect Bandwidth 	& $160$ GB/sec	\\
		\hline

  \end{tabular}
\vspace{-.2em}
  \label{tab:npu_config}
\end{table}

{\bf Benchmarks.} We study six DL applications  as part of our dense DNN
workloads.  We chose AlexNet, GoogLeNet, and
ResNet~\cite{alexnet,googlenet,resnet} as our CNN application suite (denoted as
		CNN-1/CNN-2/CNN-3, respectively) because they cover a wide range of filter
and activation sizes. We also include three RNNs from
DeepBench~\cite{deepbench}, one regular GEMV (general matrix-vector
		multiplication) based RNN (RNN-1) and two LSTM based RNNs (RNN-2/RNN-3).
For these workloads, we observe an intractable amount of simulation time when
the batch size is larger than $16$, so our analysis assumes a batch size of
$1$/$4$/$8$ (denoted as \texttt{b01/b04/b08}), which is reasonable for
inference scenarios.  To accommodate training scenarios, we experiment a subset of the layers (i.e.,
		a common layer configuration exhibited in each of our DNN) with large batch
sizes in \sect{sect:sensitivity} as a sensitivity study to explore the
implication of address translation on large batch training.  When studying the
effectiveness of \nmmu in handling sparse DNN layers (e.g., embedding layers)
	in \sect{sect:numa_npu}, we use two recommendation system models: the neural collaborative filtering (NCF)
	based recommendation system~\cite{he:www:2017} from MLPerf~\cite{mlperf} and the recently
	open-sourced DLRM (deep learning recommendation model) from Facebook~\cite{facebook_dlrm}.

{\bf Page sizes (small vs. large).} A key consideration in designing a virtual
memory system is its page size. Compared to baseline $4$KB
pages, large ($2$MB) pages can potentially reduce the translation
invoked stalls by increasing TLB reach and reducing TLB misses.
As we discuss in \sect{sect:large_pages}, we find that $2$ MB large pages do in
fact decrease the performance overhead of address translations for conventional,
dense CNNs/RNNs which exhibits highly regular dataflows. Unfortunately, for emerging DL
applications employing sparse embedding layers  with irregular memory access
patterns (\sect{sect:why_npu_mmu}), we find that demand paging with large page
sizes incurs significant performance loss compared to small pages (an average
		$83\%$ vs. $99\%$ performance loss for small vs. large pages).
Consequently, large pages alone are no silver bullet in designing a  virtual
memory system for NPUs, motivating the importance of robust address translation
for small pages.  As large pages perform well for conventional CNNs/RNNs,
		we assume the baseline $4$KB pages for our default evaluation.
		We revisit the implication of large pages on address translations, its
		pitfalls for emerging, sparse DNN layers in \sect{sect:large_pages}.

\begin{figure}[t!] \centering
\includegraphics[width=0.435\textwidth]{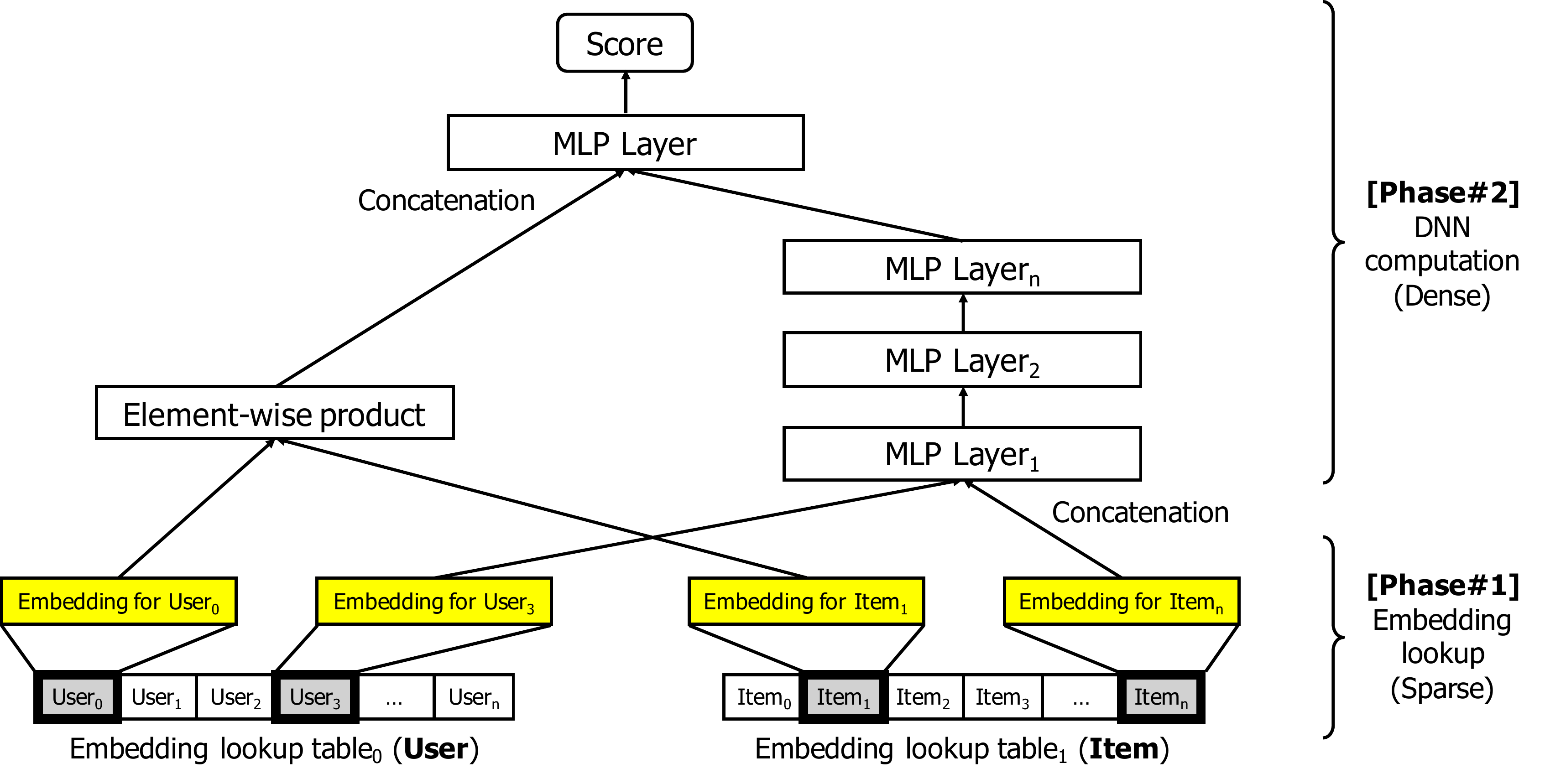}
\caption{
Neural network based collaborative filtering (NCF)~\cite{he:www:2017}. Reading out embedding vectors (e.g., the yellow colored four vectors) from the user/item embedding lookup tables  is conceptually similar to a \emph{gather} operation with very low temporal and spatial locality. 
}
\label{fig:ncf}
\end{figure}

\section{Motivation: Why MMU for NPUs?} 
\label{sect:background}

\subsection{Emerging, Memory-limited DL Workloads}
\label{sect:embedding}

A common property that conventional DL applications share (for both
		training/inference) is that its working set always fits within the tens of
GBs of NPU local memory budget -- an artifact of the physically-addressed NPU
memory.  However, recent studies from several
hyperscalars~\cite{hestness:2019:ppopp,park:2018:fb} project that emerging DL
workloads are heavily memory ``capacity'' limited, exhibiting several hundreds
of GBs of memory footprint. DL applications such as recommendation
systems~\cite{he:www:2017}, word/character language
models~\cite{jozefowicz:arxiv:2016}, and speech
recognition~\cite{chan:icassp:2016}, for instance, employ \emph{embedding
	layers} which require several tens to hundreds of GBs of memory to store just
	the model weights themselves, even for inference.  \fig{fig:ncf} illustrates
	the usage of embedding layers in recommendation systems that incorporate
	neural networks~\cite{he:www:2017}, which are the current state-of-the-art
	algorithms being deployed for news feed, search, and ads.  Facebook, for
	instance, stores deep learning features (e.g., the pages a particular user
			liked) as vectors called \emph{embeddings} which are utilized to
	recommend relevant posts or contents to users~\cite{park:2018:fb}.  Each user
	has a unique embedding vector so the total number of vectors scale
	proportional to the number of users.  Embedding layers therefore house
	billions of weight parameters, which leads to its tens to hundreds of GBs of
	memory usage.  As shown in \fig{fig:ncf}, recommender systems consist of two
	phases: 1) an embedding ``lookup'' phase that \emph{gathers} multiple
	embedding vectors from potentially multiple lookup table (e.g., two tables in
			\fig{fig:ncf}) to batch them into a single tensor, and 2) using the
			batched tensor to execute several multi-layer perceptron (MLP) layers.
			Because the  model size of these embedding lookup tables are far
				beyond the memory capacity limits of GPUs/NPUs, the solutions vendors predominantly take are:

\vspace{0.3em} 
\begin{enumerate} 

\item {\bf \emph{Host-centric}} approach: all the embedding lookup tables are stored in the
capacity-optimized CPU memory and CPU is solely used for the entire inference
process~\cite{facebook_dlrm,dlrm:arch}

\item	{\bf \emph{Accelerator-centric}} approach: model-parallelism~\cite{alex_weird_trick} is
used to partition the embedding tables across multiple GPU/NPU's
bandwidth-optimized memory~\cite{facebook_dlrm,hestness:2019:ppopp}, addressing the memory capacity constraints of
embeddings.  

\end{enumerate} 
\vspace{0.3em}

 \begin{figure}[t!] \centering
\hspace{2em}\includegraphics[width=0.33\textwidth]{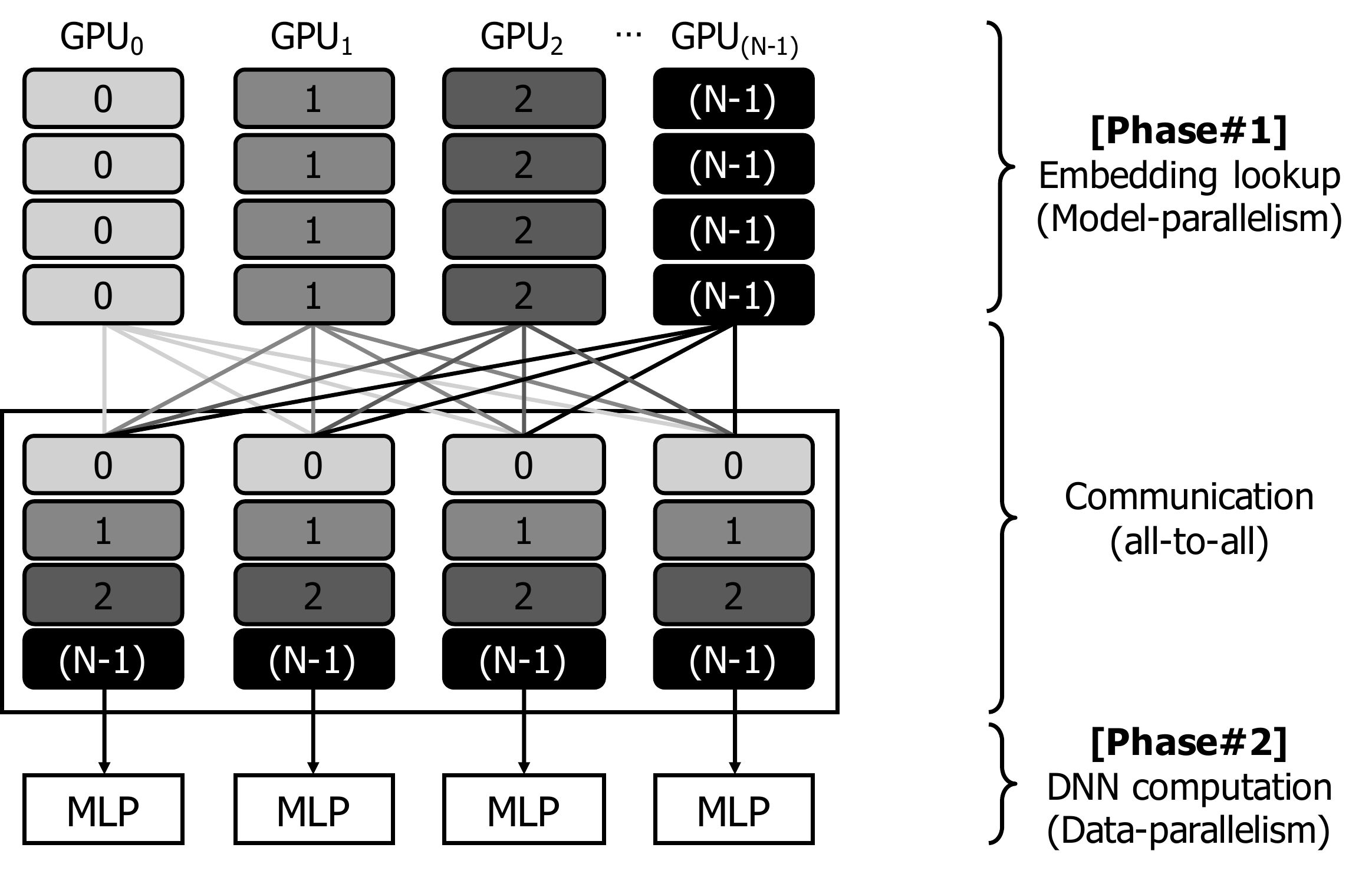}
\caption{
Facebook's accelerator-centric parallelization scheme~\cite{facebook_dlrm,hestness:2019:ppopp} as employed 
in its DLRM model.
Each GPU is allocated with $1$/$N$
	of the embedding tables, so the embedding lookup phase is model-parallelized
	(e.g., GPU$_{0}$ stores table$_{0}$,
		GPU$_{1}$ stores table$_{1}$, $\ldots$).
	As the MLP portions are parallelized using data-parallelism, each GPU must have
	all its share of embeddings ready before MLPs are executed. Consequently, 
	an all-to-all communication is conducted to gather the embeddings from all
		the neighboring GPUs, shown graphically where each color denotes a
		different element of the minibatch and each number denotes the GPU and
		the embeddings allocated to it. This figure is reproduced from an article by Facebook announcing the open-sourcing of their DLRM model~\cite{facebook_dlrm}.
}
\label{fig:parallelized_dlrm}
\end{figure}

\subsection{NPU MMU for Remote Memory Access}
\label{sect:why_npu_mmu}

 \fig{fig:parallelized_dlrm} shows a DNN-based recommendation system
 parallelized in an ``\emph{accelerator-centric}'' fashion
 (\sect{sect:embedding}).  That is, the compute-dominated MLPs are parallelized
 using data-parallelism to improve performance of MLPs, whereas the
 memory-capacity limited embedding tables are model-parallelized to overcome
 the constraints of (only tens of GBs of) accelerator local memory capacity.
 Assuming different accelerator is allocated with a different lookup table, an
 \emph{all-to-all} communication is required in order to shuffle the results of
 an embedding lookup of an entire minibatch on each accelerator into parts of a
 minibatch of embedding lookups on all accelerators. If the accelerators we
 assume here are GPUs, they have several options that enable the all-to-all
 communication process: 1) all GPUs can be passed with a (shared) pointer to
 each embedding table, potentially stored in a remote GPU's local memory, which
 allows any GPU with the pointer to directly load data in a CC-NUMA fashion
 (over NVLINK~\cite{ararwal:asplos:2015,nvlink}), or 2) use P2P
 \texttt{cudaMemcpy} to initiate direct GPU$\leftrightarrow$GPU DMA copy
 operations without having to utilize host-side pinned memory as an
 intermediate step.  Unfortunately, neither of these options are available for
 an MMU-less NPU because it does not have the ability to address memory that is
 outside its local, physical memory address space. In other words, the NPU is
 not able to reference data that is not already available within its physical
 memory.  As such, the CPU runtime must manually copy the embeddings from the
 source NPU memory to an intermediate CPU-side pinned memory, and then do
 another copy of these embeddings into the destination NPU memory.
 As we quantitatively detail in \sect{sect:numa_npu}, such multi-step data copies and data duplication	adds significant
 latency, leading to an average $71\%$ performance overhead.

				Given this landscape, we argue that NPUs are in urgent need for
				architectural support for robust address translations.  In the
				remainder of this section, we first discuss the fundamental
				architectural differences between GPUs and NPUs and the limitations of
				blindly employing prior GPU-centric MMUs as-is.  We then motivate the need of an
				NPU-optimized MMU design based on a data-driven approach. We re-visit
				the usefulness of our NPU MMU design in handling DL applications using sparse embedding
				layers in \sect{sect:numa_npu}, which improves the
				performance of an MMU-less NPU by $3.4\times$.

\subsection{Data-driven Analysis of NPU MMUs} 
	\label{sect:observation}

\begin{figure}[t!] \centering
\includegraphics[width=0.42\textwidth]{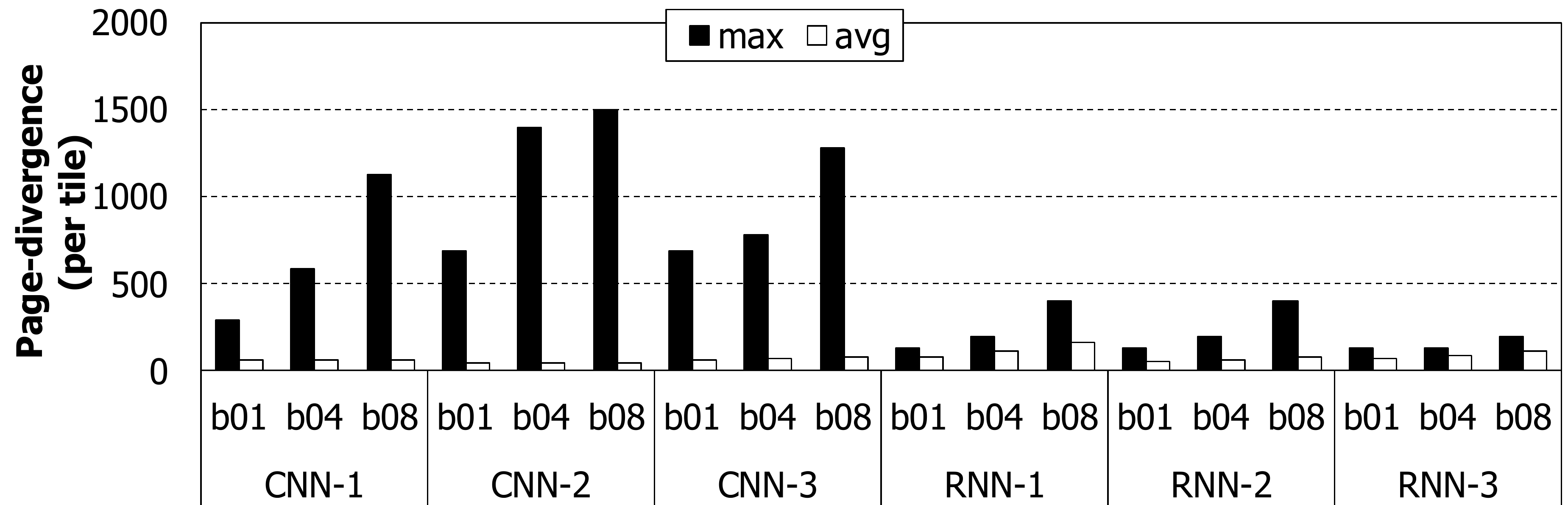}
\caption{
Maximum/average number of distinct pages accessed for each tile fetched by the DMA unit under $4$KB pages. 
}
\label{fig:avg_max_page_divg}
\end{figure}

{\bf Translation bursts in SPM-centric NPUs.} As discussed in
\sect{sect:npu_arch}, the data movements between main memory and SPM are
conducted in coarse-grained tile chunks, which can be several
MBs.  For instance, our baseline NPU employs $10$ MB of
SPM each for \ia and \weights, 
	so the tile size of \ia and \weights can be as large as
	($10$/$2$)=$5$ MB.  Putting this number into perspective, assuming
	NPUs have an MMU that enables VA-to-PA translations, a single tile request
	by the DMA seeking to fully populate the $5$ MB on-chip SPM will
	need to access a minimum of ($5$ MB/$4$ KB) = $1.2$K distinct
	pages under the baseline 4 KB page. 
	The actual number of pages accessed can be
	much larger than this minimum number because the DMA is not necessarily
	fetching data in page-granularity in a dense fashion (i.e., worst case, the
		DMA fetches only a single word from a single page in a sparse manner).
	\fig{fig:avg_max_page_divg} illustrates the average and maximum number of
	distinct pages accessed  by a single tile requested by the DMA.

\begin{figure}[t!] \centering
\subfloat[]{
	\includegraphics[width=0.45\textwidth]{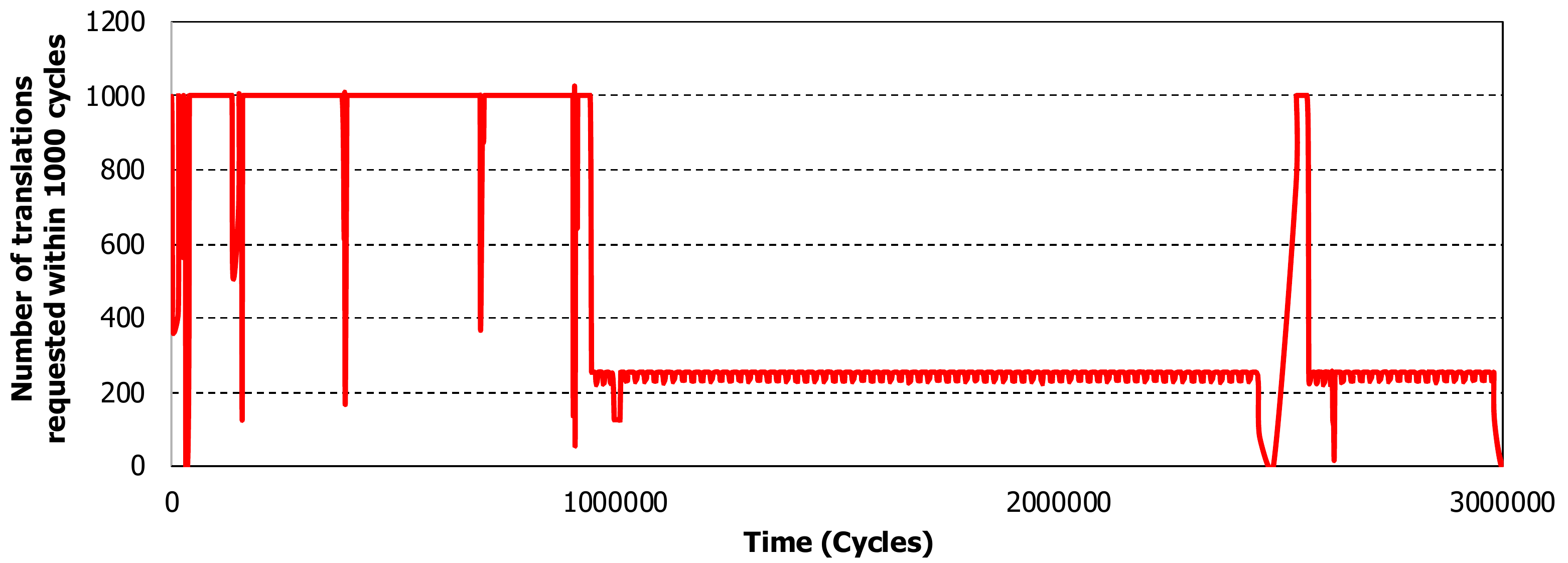}
	\label{fig:translation_bursts_cnn}
}
\vspace{0em}
\subfloat[]{
	\includegraphics[width=0.45\textwidth]{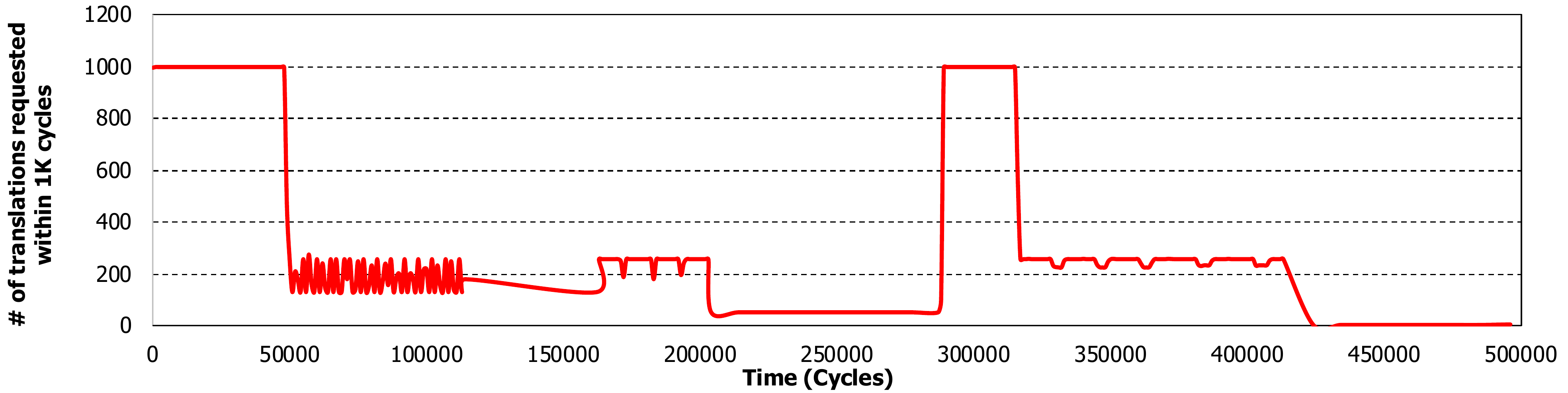}
	\label{fig:translation_bursts_rnn}
}
\vspace{0em}
\caption{
Number of address translations requested by the DMA assuming $4$ KB pages within a consecutive $1000$ clock cycle window ((a) CNN-1 and (b) RNN-1). 
	The DMA unit sends a single translation each cycle, so $1000$ on the y-axis represent phases where the DMA generates a burst
of translations.
}
\label{fig:translation_bursts}
\end{figure}

		Note that the
		\ia/\weights tiles are multi-dimensional tensors	mapped to a traditional, 
		linear (1D) DRAM memory. Consequently, a single tile tensor can 
		be decomposed into multiple, linearized memory transactions by the DMA
		unit. Each of these memory transactions require address translation to
		determine which page it belongs to, so the actual number of translations
		invoked can be much larger than the number of pages accessed
		(\fig{fig:avg_max_page_divg}).  To make matters worse, these address
		translation requests are generated in large {\bf bursts} within a short
		timeframe (henceforth referred to as \emph{translation bursts}), which
		cause significant translation bandwidth pressure on the MMU
		(\fig{fig:translation_bursts}).  \old{Not surprisingly, large pages help reduce
		the number of different pages accessed per each tile request (e.g.,
				$125\times$ reduction for CNN-1).  Nonetheless, the number of memory
		transactions that require address translations are still high as this is a
		function of how the multi-dimensional tensor tile is mapped into the linear
		DRAM system. For instance, while the number of distinct pages accessed per each tile request in CNN-1
		was reduced by $125\times$, the reduction in translation requests was \emph{only}
		$4\times$ when using $2$ MB pages as opposed to $4$ KB pages.}  While these
		numbers might at first glance seem surprising, we observe that this is a
		natural outcome of NPU architectures optimized for data-/task-level
		parallelism using an SPM based on-chip memory hierarchy.  State-of-the-art
		NPUs typically contain tens of thousands of ALUs on-chip, so the SPM must
		be large enough to seamlessly feed these processing engines with useful
		work.  As there is an implicit \emph{barrier} enforced at the boundaries of
		a particular tile's compute and memory phase (i.e., any given
				\texttt{tile$_{(n)}$}'s computation can be initiated only when the
				entire tile is fully fetched into the SPM, see \fig{fig:tiling_timeline}),
		the DMA unit tries to concurrently launch the data read requests to DRAM to
		maximize memory-level parallelism and fetch \ia/\weights tiles as soon as
		possible, inevitably leading to translation bursts.

{\bf Pitfalls of GPU-centric MMUs for NPUs.}  As convolutions or matrix-multiplication operations 
are well-known to exhibit high data reuse thanks to its regular dataflow~\cite{eyeriss}, 
		one might think that conventional TLBs should effectively capture
the translation reuse with high TLB hit rates. However, NPUs have fundamental architectural
differences than GPUs, rendering prior GPU-centric MMU solutions ineffective in handling the
aforementioned translation bursts. Recall that state-of-the-art NPU architectures are based on
a SPM-centric memory hierarchy. The
(PE$\leftrightarrow$SPM) data traffic do not require address translations
because SPM is addressed using VAs rather than PAs. Consequently, unlike a GPU
where a per-core, post-coalescing TLB can effectively reduce a substantial amount
of GPU translation requests~\cite{gpu_tlb,gpu_x86_at}, a per-PE TLB cannot
	help in fitering out the translation burst bandwidth pressure of NPUs.
	However, the intra-tile translation locality does exist for
	(SPM$\leftrightarrow$DRAM) traffics when the DMA unit invokes multiple data
	fetch requests from memory to the SPM that fall under the same page.  
	We observe however that such translation locality is not adequately
	captured with a conventional TLB hierarchy because the bursts of translations often
	query the TLB even before the PTW delivers the VA-to-PA translations! 
	Such phenomenon is a unique characteristic of the SPM-centric NPUs: for GPUs, memory
	read/write operations are initiated through load and store instructions,
	which only amounts to $10$$-$$20\%$ of the instruction mixes~\cite{leng:isca:2013} and is therefore
	likely to be distanced apart in time when sent over to the MMU for address
	translations. The SPM-centric NPU however invokes bursts of these translation
	request traffic to the MMU within a short time-window leading to its high translation throughput
	requirements.

\begin{figure}[t!] \centering
\includegraphics[width=0.47\textwidth]{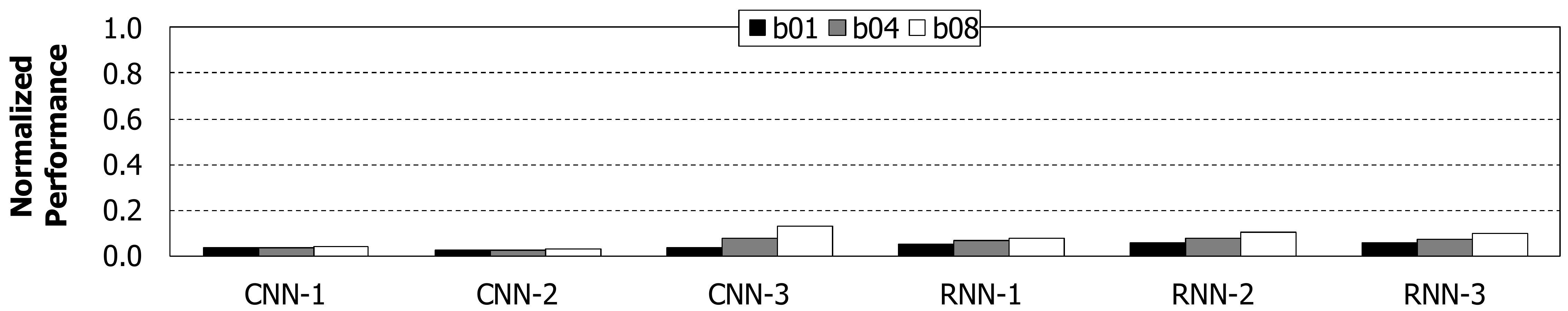}
\caption{
Normalized performance when enabling address translation in $4$ KB page granularity. The baseline IOMMU design
which contains 2048 TLB entries and 8 hardware page-table walkers were used.
Results are normalized to an oracular MMU that assumes all translations hit in the TLB with no additional TLB access latency.
}
\label{fig:iommu_naive_overhead}
\end{figure}

{\bf Translation throughput vs. translation locality.} In general, we observe
that NPUs that utilize a naive, strawman IOMMU design enhanced with some key
GPU-centric optimizations as-is (i.e., per-PE TLBs, parallel PTWs, a local
		multi-level TLB hierarchy) is not able to properly handle the NPU
translation bursts as it is \emph{optimized for translation locality rather
	than translation throughput}, experiencing severe performance slowdown as
	shown in \fig{fig:iommu_naive_overhead}.   In effect, the sheer volume of
	translations requested to the IOMMU leads to a large number of page-table
	walks, even after being filtered by the TLBs.  These massive number of
	address translations eventually becomes bottlenecked by the limited
	parallelism provided with a handful of (eight) shared IOMMU PTWs.  Overall,
	our data-driven analysis shows that conventional MMUs which are primarily
	designed to capture translation locality (i.e., TLBs), rather than
	translation throughput (i.e., number of PTWs), are inadequate in handling the
	translation bursts in NPUs.  To validate whether the TLB can be a primary
	target for improvement in NPU MMUs, we sweep the numbers of TLB entries on
	top of our baseline IOMMU with eight hardware PTWs. Even with an
	unrealistically large TLB with $128$K entries ($64\times$ increase over
			baseline $2$K TLB entries, \tab{tab:npu_config}), the NPU fails to
	completely filter out the bursts of translation requests, achieving less than
	$0.02\%$ performance improvement than baseline IOMMU.
Overall, we conclude that NPU local TLBs, while beneficial, is not sufficiently
performant enough to filter out most of the translations. This is because the
bursts of address translations cause significant number of page-table walks
instantiated and be bottlenecked by the translation throughput provided with
IOMMUs.  This is in stark contrast with GPUs where prior
work~\cite{gpu_tlb,gpu_x86_at} has shown TLBs to be effective in capturing an
average 70$-$80$\%$ of translations.  One might think that by having the DMA
unit send data requests in a less bursty fashion (e.g., only allow up to a
		limited number of data and address translations that the IOMMU can
		sustain), the effectiveness of TLBs can be restored and performance loss
reduced.  Unfortunately, such design decision will inevitably reduce
memory-level parallelism and memory bandwidth utilization, significantly
slowing down memory-limited applications like RNNs.

{\bf Proposed approach: throughput-centric MMU.} Based on our data-driven
analysis, we conclude that translation throughput should be the primary design
target for NPU MMUs.  This is because of the SPM-centric NPU's unique
architectural characteristic, where the tile-based bulk DMA transfers invoke
translation bursts which are not adequately captured using the
locality-optimized, GPU-centric MMUs.  In the following section, we
propose  a ``throughput''-centric NPU MMU design that effectively balances
translation throughput while also adequately capturing translation locality.

\section{NeuMMU: Designing an MMU for NPUs}
\label{sect:nmmu_proposal}

\subsection{PRMB: Translation Bandwidth Filter} 
\label{sect:opt2}

\begin{figure}[t!] \centering
\includegraphics[width=0.44\textwidth]{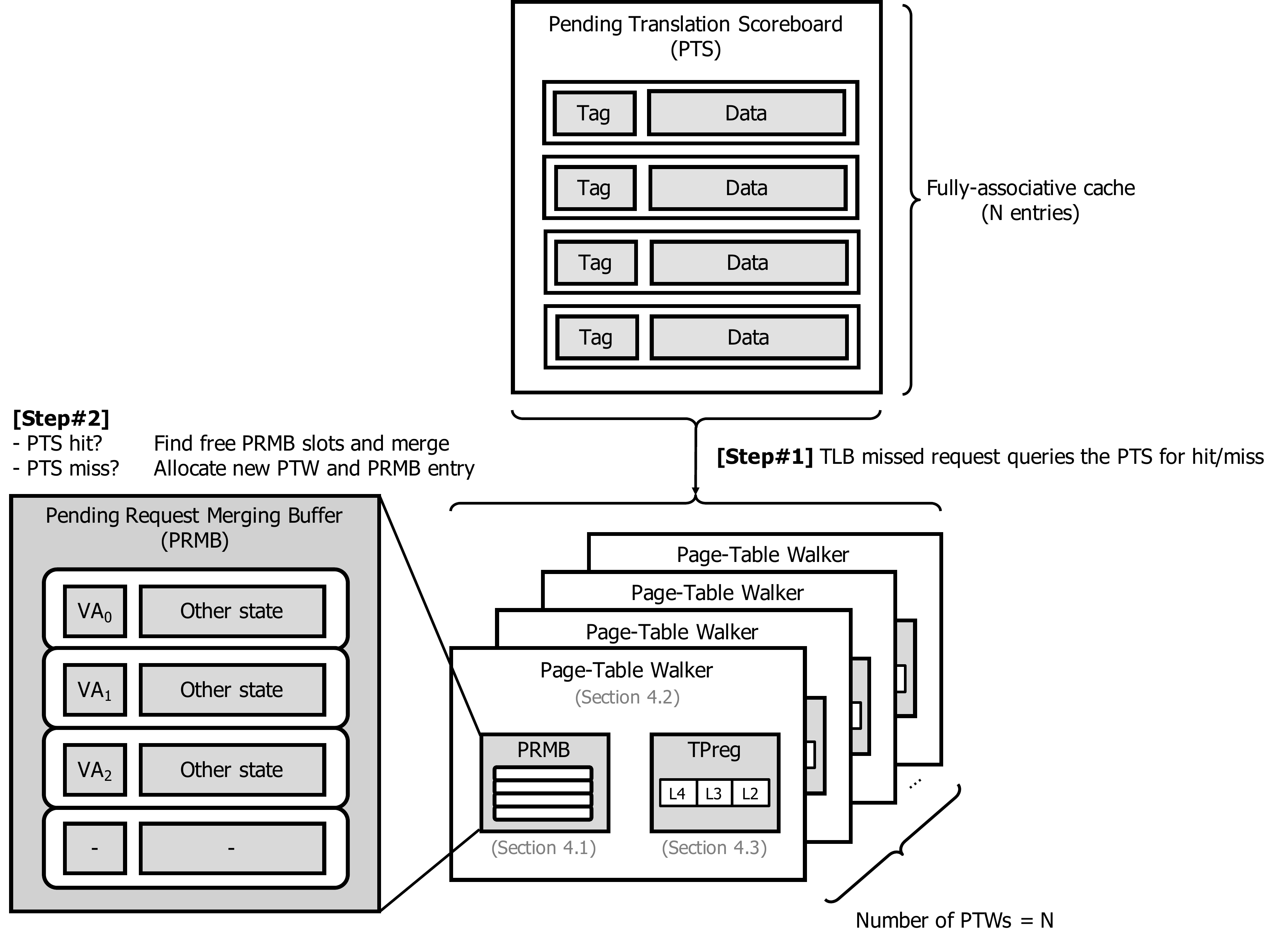}
\caption{
Proposed PTW with a \prmb containing $4$ mergeable slots. The example
assumes that the PTW is currently walking the page-tables to translate a virtual page number of \texttt{VPN$_{n}$}.
Three translation requests to the same virtual page are merged inside the \prmb.
}
\label{fig:mshr_per_ptw}
\end{figure}

As discussed in \sect{sect:observation}, a key challenge with NPU address
translation is that the SPM-centric memory hierarchy invokes a burst of several
thousands of address translations when moving data in/out of main memory
from/to SPM.  Our first proposal is based on the key observation that a
significant fraction of translation bursts hit in the same page that is already
being translated by the PTW.  To capture such translation locality within
translation bursts, we propose a PTW design enhanced with a \emph{pending request
	merging buffer} (\prmb) that \emph{absorbs} the page translation requests
	falling under an already inflight, pending translation initiated by that same
	PTW.  \fig{fig:mshr_per_ptw} illustrates the microarchitecture of the  
	proposed PTW with our \prmb assuming that it can merge up to
	$4$ identical page translations per each PTW. Any memory
	transaction that misses in the TLB is first routed to the \emph{pending
		translation scoreboard} (\pts) to check whether any one of the $N$ parallel
		PTWs is currently under the process of translating the corresponding page.
		The \pts is a fully-associative cache with $N$ cache entries (equivalent to
				the number of PTWs) and is tagged with the virtual page number (VPN). A
		hit in the \pts implies that a VA$\rightarrow$PA translation for this
		particular VPN is currently inflight. If there are vacant \prmb mergeable
		slots within the PTW, the \pts-hit request is merged inside the \prmb and waits
		until the translation comes back.  A \pts miss however implies that
		neither the TLB nor any one of the PTWs contains the translation for this
		VPN. The \pts therefore assigns one of the vacant PTWs (if any) as the designated
		translation unit to walk the page-tables, and registers the VPN information
		inside one of the \pts entries so that future translations to this particular VPN can be
		merged. When all the PTWs as well as all possible \prmb mergeable slots are full,
		any further translation requests are blocked until the translation bandwidth is available.
		
\begin{figure}[t!] \centering
\includegraphics[width=0.475\textwidth]{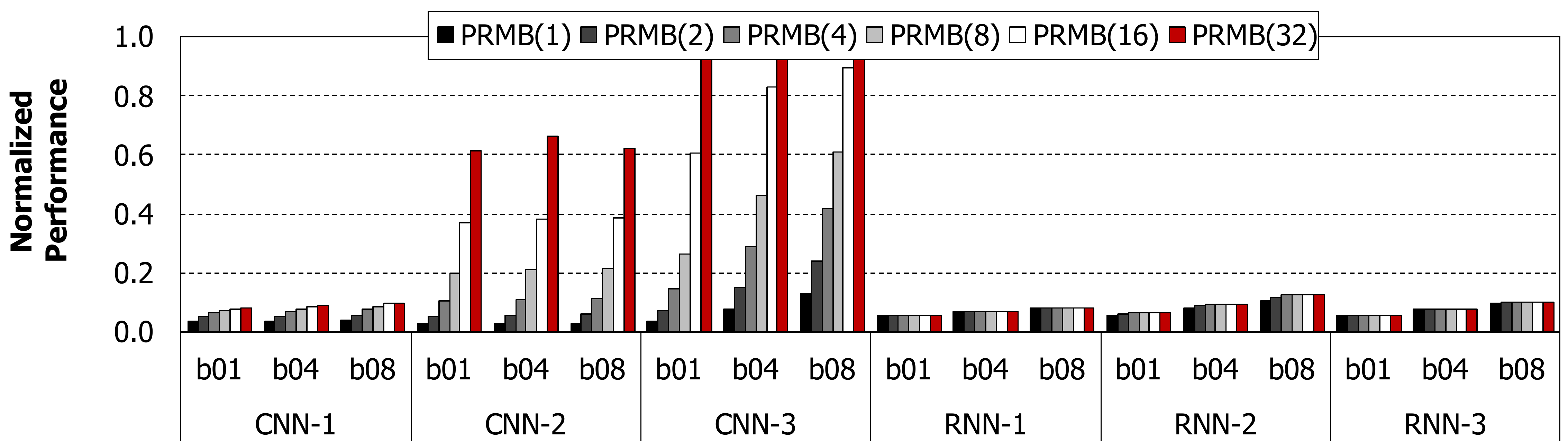}
\caption{
Performance sensitivity to the number of mergeable slots within the \prmb.
All configurations assume the baseline setting with $8$ PTWs and $2048$ TLB entries as assumed in \fig{fig:iommu_naive_overhead}.
}
\vspace{-0.2em}
\label{fig:perf_sensitivity_mshr_entries}
\end{figure}

		Because the translations that are merged inside \prmb do not send
		a separate page-walk request and instead \emph{waits} for the already
		inflight translation request to come back, our \prmb microarchitecture saves not only memory bandwidth
		but more importantly, the PTW ``translation bandwidth'', effectively
		functioning as a \emph{translation bandwidth filtering} mechanism.  Once the
		translation is available, the PTW controller queries the \prmb and returns
		the merged requests back to the DMA unit on a cycle-by-cycle basis.
		\fig{fig:perf_sensitivity_mshr_entries} shows the effect of \prmb on
		overall performance as a function of how many merge-able entries are
		provisioned inside each \prmb.  As depicted, for our studied DNNs, having
		$8$-$32$ mergeable slots per each PTW can significantly capture the
		translation burst locality thereby minimizing the redundant translation
		requests from wasting memory and translation bandwidth.  This allows our
		\prmb-enhanced NPU to achieve an average $11\%$ (max $98\%$) performance of an
		oracular MMU, a significant improvement over the
		baseline IOMMU. Nonetheless, there still exists a significant performance
		gap of $89\%$  motivating us to our second proposition.

\subsection{(Translation) Throughput-centric MMU}
\label{sect:opt3}

 The \prmb-enhanced PTW helps capture translation burst locality while
 minimizing waste in memory and translation throughput. Nonetheless, the low
 average TLB hit rate and the sheer volume of required address translations
 render significant pressure on the meager $8$ IOMMU page-table walkers.  While
 Powers et al.~\cite{gpu_x86_at} similarly observed that enhancing parallelism
 to the PTW helps improve GPU's translation throughput, adding more parallel
 PTWs was only able to achieve, on average, $30\%$ of the oracular MMU design
 point under the GPU context.  This is because leveraging translation locality,
 using the per-core/post-coalescing TLB and multi-level TLB hierarchy, was
 shown to be more important than enhancing raw translation throughput for
 GPUs~\cite{gpu_tlb,gpu_x86_at}.

\begin{figure}[t!] \centering
\includegraphics[width=0.475\textwidth]{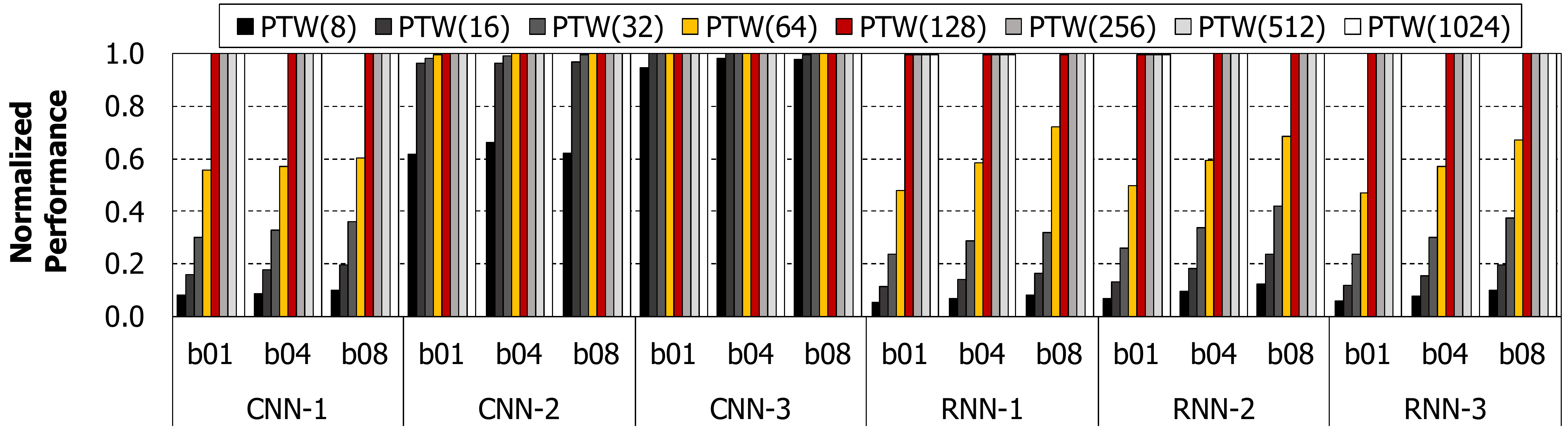}
\caption{
	Performance sensitivity to the number of hardware page-table walkers with $4$
		KB pages.   All configurations assumes a 2048 entry TLB with $32$
		mergeable slots per each \prmb unit  within each PTW.  Results are
		normalized against an oracular MMU.
}
\label{fig:perf_sensitivity_num_ptw}
\end{figure}

	 Our work, on the other hand, makes the unique observation that the bursty
	 nature of NPU translation requests, coupled with the relatively low TLB hit
	 rates, ``mandates'' a throughput-centric MMU as a primary design objective.
	 As such, the key insight our data-driven analysis delivers is that the
	 SPM-centric \emph{NPUs should be designed for
	 improving translation throughput first, and translation locality second}.  As such, our
	 second proposition is that the NPU MMU should be further enhanced for 
	 translation throughput by adding a larger number of PTWs.
 \fig{fig:perf_sensitivity_num_ptw} shows the
 NPU performance sensitivity on address translation throughput, where
 increasing the number of PTWs from $8$ to $128$ closes the performance
 gap from an average $11\%$ to $99\%$ for baseline $4$ KB pages: 
	 $128$ PTWs turned out to be a good design point for the set of
	 benchmarks we have evaluated, but larger/smaller PTWs might be required for
	 alternative NPU configurations. We discuss the sensitivity of \nmmu for
	 alternative design points in \sect{sect:sensitivity}. 
	 As noted in \sect{sect:observation}, the
	 IOMMU is designed to be shared by \emph{multiple} accelerators. To make sure
	 the NPU alone does not saturate the address translation throughput, we argue
	 that the number of PTWs be sufficiently provisioned such that it does not
	 become a performance hotspot.  Studying efficient MMU resource allocation
	 strategies across multiple accelerators for QoS is beyond our scope
	and we leave it as future work.	 

\begin{figure}[t!] \centering
\subfloat[]{
\includegraphics[width=0.475\textwidth]{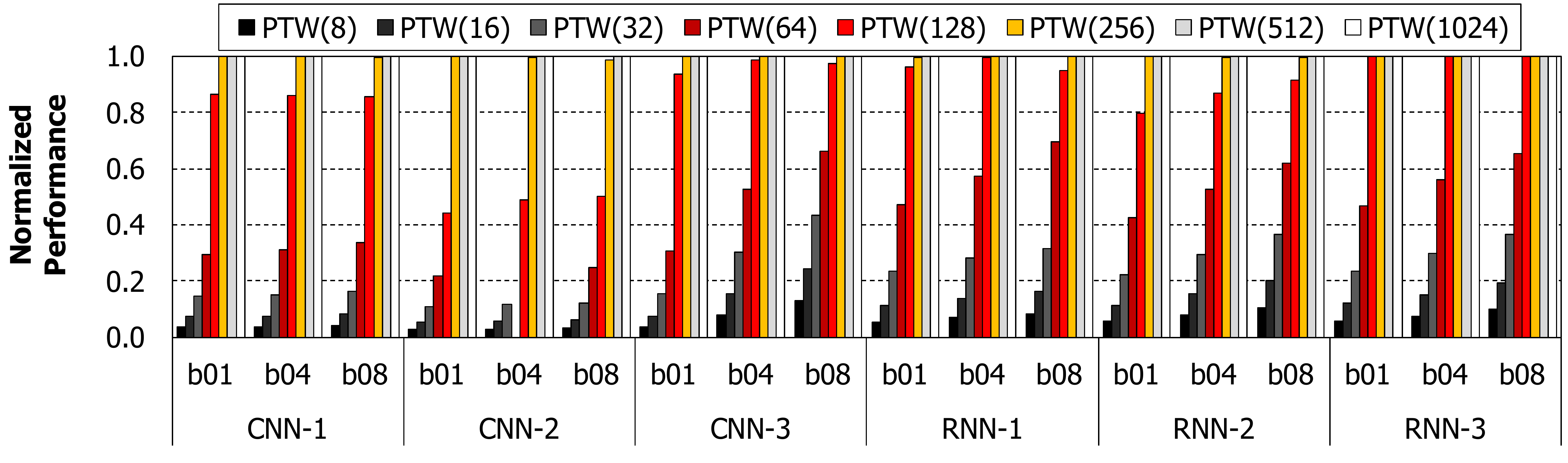}
}
\vspace{-0.5em}
\subfloat[]{
\includegraphics[width=0.475\textwidth]{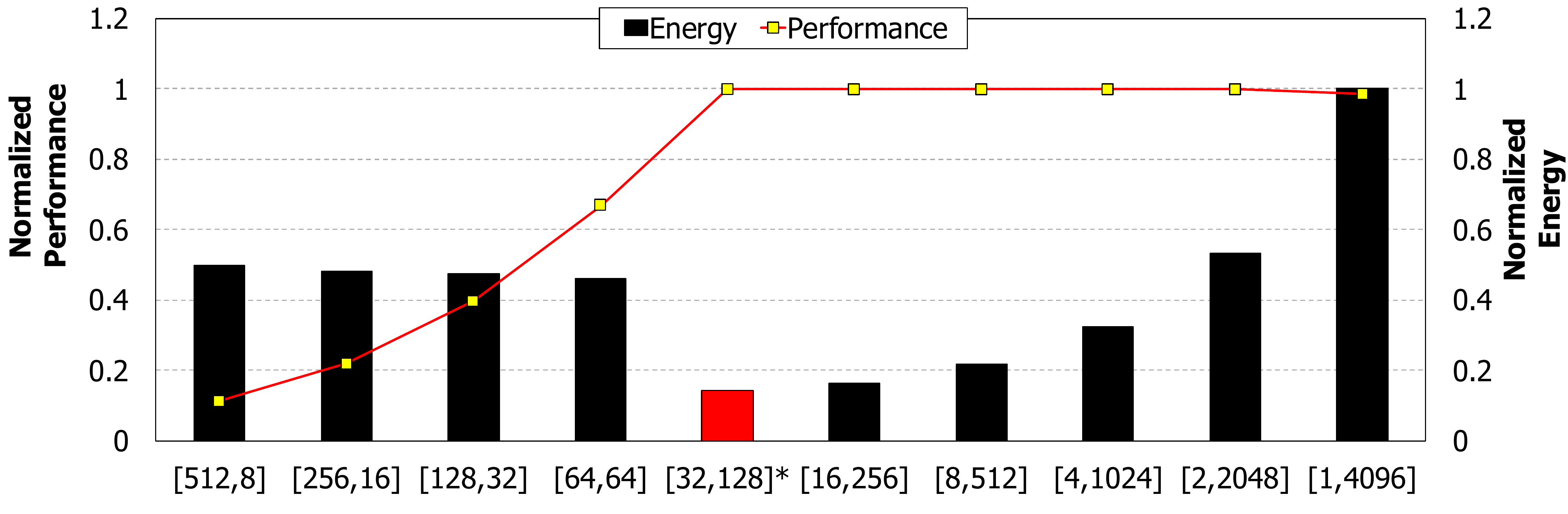}
}
\vspace{-0.5em}
\caption{
(a) Performance sensitivity to the number of hardware page-table walkers 
``without'' the \prmb microarchitecture employed. (b) Energy consumed for address translations 
when \nmmu employs \texttt{M} \prmb entries and \texttt{N} PTWs, denoted as [\texttt{M},\texttt{N}] in the x-axis.
We used the 1) energy table for a $45$ nm CMOS process~\cite{energy_table:horowitz} to model the DRAM access energy
consumption in page-table walks and 2) CACTI~\cite{cacti} for modeling \prmb access energy.
}
\label{fig:perf_sensitivity_num_ptw_without_prmb}
\end{figure}

	It is worth pointing out that blindly increasing the number of PTWs alone, \emph{without} employing 
	our translation bandwidth filtering \prmb microarchitecture, can cause significant overheads in 
	energy-efficiency.  \fig{fig:perf_sensitivity_num_ptw_without_prmb}(a) shows
	the performance of baseline IOMMU enhanced with a larger number of PTWs without our
	\prmb design adopted. With $1024$ PTWs with no \prmb, the performance does in fact match
	the performance of \nmmu with $32$ \prmb and $128$ PTWs. Such design point however consumes significantly
	more energy as shown in \fig{fig:perf_sensitivity_num_ptw_without_prmb}(b). Without the translation bandwidth
	filtering effects of \prmb, a significant fraction of translations that walk the page-tables are \emph{redundant} and
	causes up to $7.1\times$ more energy consumption than the nominal $32$ \prmb and $128$ PTWs of \nmmu.
	Our novel \prmb microarchitecture and the throughput-centric parallel PTW design effectively balances
	performance and energy-efficiency, reaching $99\%$ of the performance of oracle while consuming
	 much less energy than single-handedly relying on large PTWs without \prmb.

\subsection{Translation Path ``Registers'' (not Caches)}
\label{sect:opt4}

An important challenge with the \nmmu design so far is that a significant
fraction of translations still require a page-table walk.  While the abundant
address translation throughput provided with \nmmu design so far effectively hides the
latency of translations, the number of page-table walks themselves are still
relatively high because of the low TLB hit rate. As our study assumes an x86-64
style, hierarchical 4-level page-tables, a single page-table walk operation
would incur up to four memory transactions with significant power overheads. For
power-limited environment, the overhead of adding address translations can be 
prohibitive which leads to our last proposal: a lightweight
\emph{translation path ``register''} (\tpr) that allows PTWs to \emph{skip}
some page-table walking steps. Our \tpr microarchitecture is inspired
by the well-known MMU caches~\cite{mmu_cache} (aka translation path caches), widely adopted in CPUs/GPUs, but
\tpr leverages the unique characteristics of DNNs to minimize its 
implementation overheads (i.e., less than $16$ bytes per PTW) while
reducing the number page-table walk invoked memory transactions by more than $2.5\times$.

{\bf Benefits of caching translation paths.} Under x86-64 based translation system, the paged
virtual memory is implemented using a radix tree for their page-tables.  The
translation path caches accelerate the page-table walking process by allowing
the processor (in our case, the NPU PTW) to skip over a single or more levels
of the radix tree. 
The virtual address
space is decomposed into a page-number and a page-offset, where the page number
is further split into a sequence of indices, four in x86-64. The first index
(L4) is used to select an entry from the root of the radix tree, which could
potentially contain a pointer to a node in the next lower level (i.e., L3) of
the tree. If a valid entry is found, the next index value is used to jump
to the next tree level, which can again potentially find a pointer to the node
in the next lower level (L2) of the tree. Such procedure is repeated
until the selected entry is invalid or the tree search finalizes at a data page
using the PA. As x86-64 currently uses $48$ bits out of the memory addressable
$64$ bits, a baseline $4$ KB page size utilizes the lower $12$ bits and the
remaining $36$ bits are divided into four $9$ bit indices.
Because page-table walks to two consecutive VA pages will most likely use the
\emph{same} L4/L3/L2 entries, significant translation locality exists across spatially
close VA regions.

\begin{figure}[t!] \centering
\includegraphics[width=0.485\textwidth]{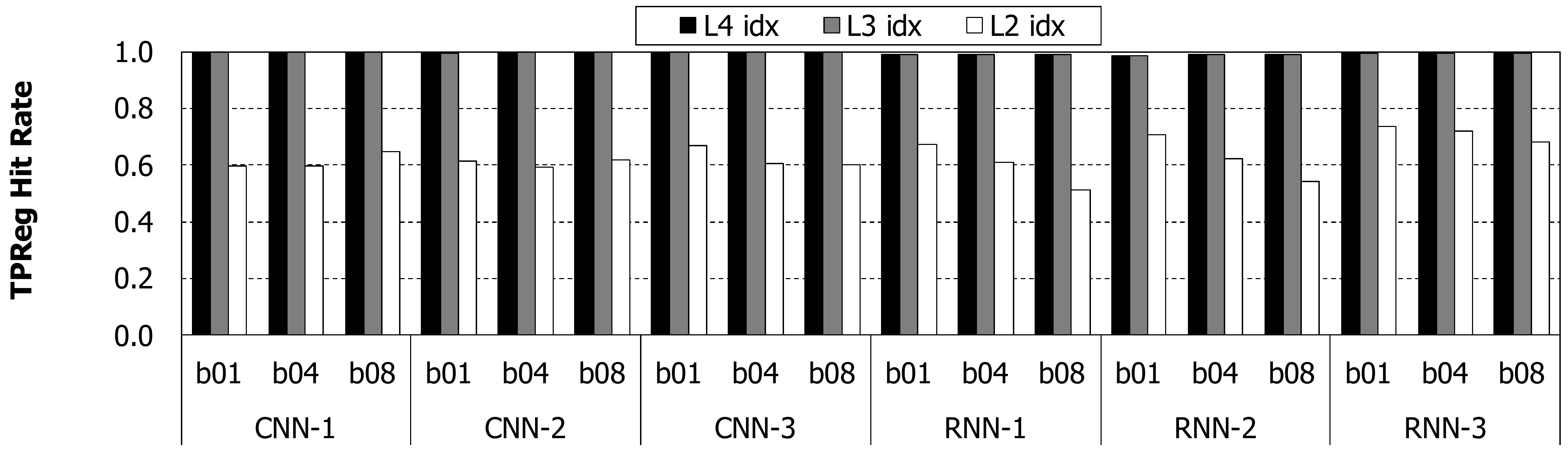}
\caption{
 TPC tag matching rate at L4/L3/L2 indices when only a single TPC entry is used (i.e., \tpr).
}
\vspace{-0.2em}
\label{fig:perf_sensitivity_tpc}
\end{figure}

{\bf Design space of translation path caches.} x86-64 processor vendors already
employ private, low-latency translation path caches that store upper-levels of the
page-table entries~\cite{mmu_cache:intel,mmu_cache:amd} and the tradeoffs of
alternative translation path cache designs are well-understood through prior
literature~\cite{mmu_cache}. A full design space exploration of all available
options for our NPU MMU design is beyond the scope of this work. Nevertheless,
				we briefly discuss two representative design points that are inspired
				by translation path caches employed by CPUs from Intel/AMD, which drives our
				proposed \tpr design.  The most intuitive implementation of a
				translation path cache is to store page-table entries tagged by the
				corresponding entry's \emph{physical address} in memory. Entries from
				different  levels of the (L4/L3/L2/L1) page-tables are mixed and shared
				inside a unified cache, all indexed and tagged by their physical
				address. Such \emph{unified page-table cache} (UPTC) is known to be
				adopted in AMD's processor designs. Intel, on
				the other hand, employs a translation cache design that is tagged using
				the virtual address. The \emph{translation-path cache} (TPC)
	microarchitecture~\cite{mmu_cache}, for instance, is tagged by the L4/L3/L2
	indices of the \emph{virtual address}. Key intuition behind the TPC design is
	that the three separate UPTC cache entries allocated to keep track of the
	three page-table lookups can be merged into a single TPC lookup when: 1) all
	physical page numbers are concatenated and merged into a single data
	entry, and 2) is tagged using a concatenation of the virtual L4/L3/L2
	indices. Under such design, a single TPC entry corresponds to an entire path,
	including all of the intermediate entries for a given page-table walk
	operation.  

	Our study reveals that TPC is much more effective than UPTC in
capturing the NPU address translation locality. On average, the L4/L3/L2 tag hit rate of
TPC was $99.5\%$/$99.5\%$/$63.1\%$ across the studied
workloads  whereas UPTC achieved an average $92.4\%$ hit rate. This allows
TPC to reduce $59\%$ less page table walks when compared to UPTC.

{\bf Translation path ``registers'' (not caches).} Based on our design space
exploration above, we conclude that a TPC-based translation caching to be a
more robust architecture than UPTC.  As shown in
\fig{fig:perf_sensitivity_tpc}, employing	a single translation path
``register'' (\tpr) per each PTW (which caches the L4/L3/L2 entries as done in
		TPC) can capture most of the performance benefits of translation caching
while removing significant fraction of the redundant page-table walks.  As
such, \tpr can be a lightweight, cost-effective solution to reduce the number
of memory transactions for NPU page-table walks.  Below we detail the {\bf key
	insights} behind the effectiveness of our \tpr microarchitecture.

	\begin{figure}[t!] \centering
\includegraphics[width=0.40\textwidth]{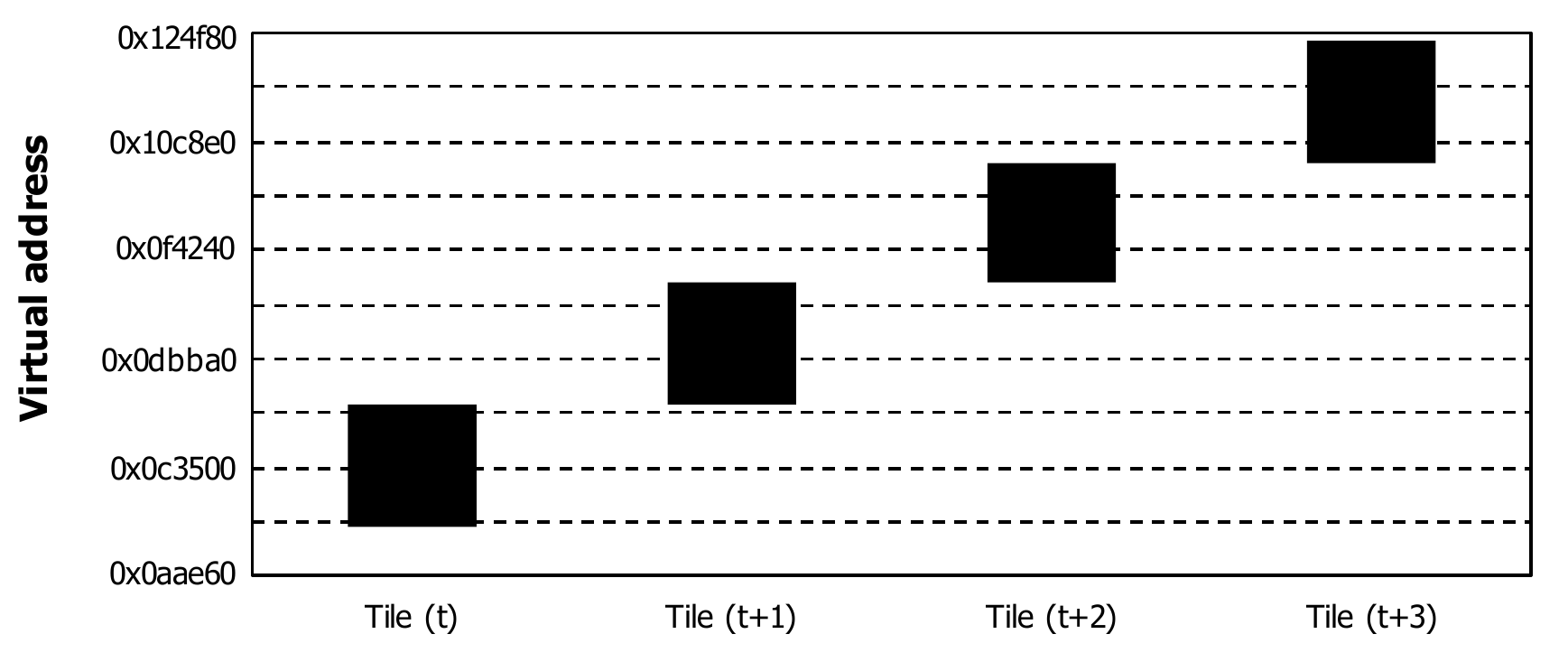}
\vspace{-0.5em}
\caption{
Trace of AlexNet's virtual address regions accessed while consecutive tiles are requested by the DMA unit.
}
\label{fig:alexnet_streaming}
\end{figure}

\begin{enumerate}
\item The \emph{number of distinct VA regions} accessed is confined within a
handful of large segments in the VA space (i.e., \ia and \weights), so translations to VA pages that fall under the 
same (\ia/\weights) segment are highly likely to share a common L4/L3/L2 translation
entry. 

\item Another important observation we make is that the DMA unit initiates tile fetch 
requests for \ia and \weights one at a time, meaning the data fetch request is not interleaved 
across \ia and \weights. This implies that the majority of address translation requests invoked
in the memory-to-SPM data fetch process will naturally share the L4/L3/L2 entries. \fig{fig:alexnet_streaming} 
illustrates the virtual addresses accessed in time, confirming the high temporal locality of address translations
which translation caching can effectively take advantage of. 

\item While the upper L4/L3/L2 entries exhibit high temporal
locality, the locality in the lower entries can be low because 
the VA accessed in time exhibit a \emph{streaming} access pattern as exhibited
in \fig{fig:alexnet_streaming}. This allows a TPC-style
translation cache with only a handful of entries to be able to capture most of the performance
benefits of translation path caches (\fig{fig:perf_sensitivity_tpc}), motivating our lightweight 
translation path ``register'', not a full-blown
cache microarchitecture.

\end{enumerate}

{\bf Energy-efficiency improvements.} While the effect of \tpr
on performance is small, its impact on
energy-efficiency is substantial.  Using the energy table for a $45$ nm
CMOS process~\cite{energy_table:horowitz}, we derive the energy overheads of
walking the page-tables for the two design points: our \nmmu with 
$128$ PTWs/$32$ \prmb entries with and without the single entry \tpr.
Our lightweight \tpr substantially
reduces the energy-overheads by an average $2.7\times$ thanks to the high
translation hit rates and the resulting smaller number of memory transactions to
walk the page-tables.

\subsection{Putting Everything Together}
\label{sect:ultimate_design}

Overall, we show that the baseline IOMMU 
fails to capture the translation locality and
throughput requirements of NPU MMUs, causing an average $95\%$ performance
overhead for baseline pages.  Based on a data-driven application
characterization study, we motivated the need for a throughput-centric MMU and
proposed three unique solutions tailored for the algorithmic nature of DNNs and
the SPM-centric NPU architecture. Putting all three solutions together, our
\nmmu design incurs an average $0.06\%$ performance overhead for baseline/small
pages when compared against an oracular MMU that assumes all translations hit
in the TLB with no additional TLB access latency.  Furthermore, \nmmu consumes
$16.3\times$ less energy than the baseline IOMMU, thanks to the \prmb and \tpr,
	which reduce the number page-table walk invoked memory transactions by
$18.8\times$.

\subsection{Implementation Overhead}
\label{sect:overhead}

We measure \nmmu's design overhead using CACTI and synthesized implementations over an
FPGA board.  The additional SRAM storage required for the per-PTW
\prmb and \tpr is as follows.  Each \prmb entry is conservatively estimated to
be $8$ bytes, so a total of ($8$$\times$$32$$\times$$128$) = $32$KB SRAM
storage is needed across the $128$ PTWs, $32$ \prmb entries per each PTW.
Regarding the \tpr, each consumes $16$ bytes so the $128$ PTWs consume $2$KB.
The \pts is a fully-associative cache with $128$ cache entries, each entry
sized at $6$ bytes.  All these amount to an area of $0.10$ mm$^2$
under $32$ nm with $13.65$ mW of leakage power consumption when estimated with
CACTI $6.5$~\cite{cacti}.  We also synthesize both the baseline
IOMMU and \nmmu on a Xilinx Virtex UltraScale+ VCU1525 
dev board and compare its resource usage. 
The amount of additional resources \nmmu consumes are less than $0.01\%$ of the
available resources, incurring negligible overheads.

\section{Case Study: NUMA NPUs for Sparse Embedding Layers}
\label{sect:numa_npu}

\begin{figure}[t!] \centering
\includegraphics[width=0.45\textwidth]{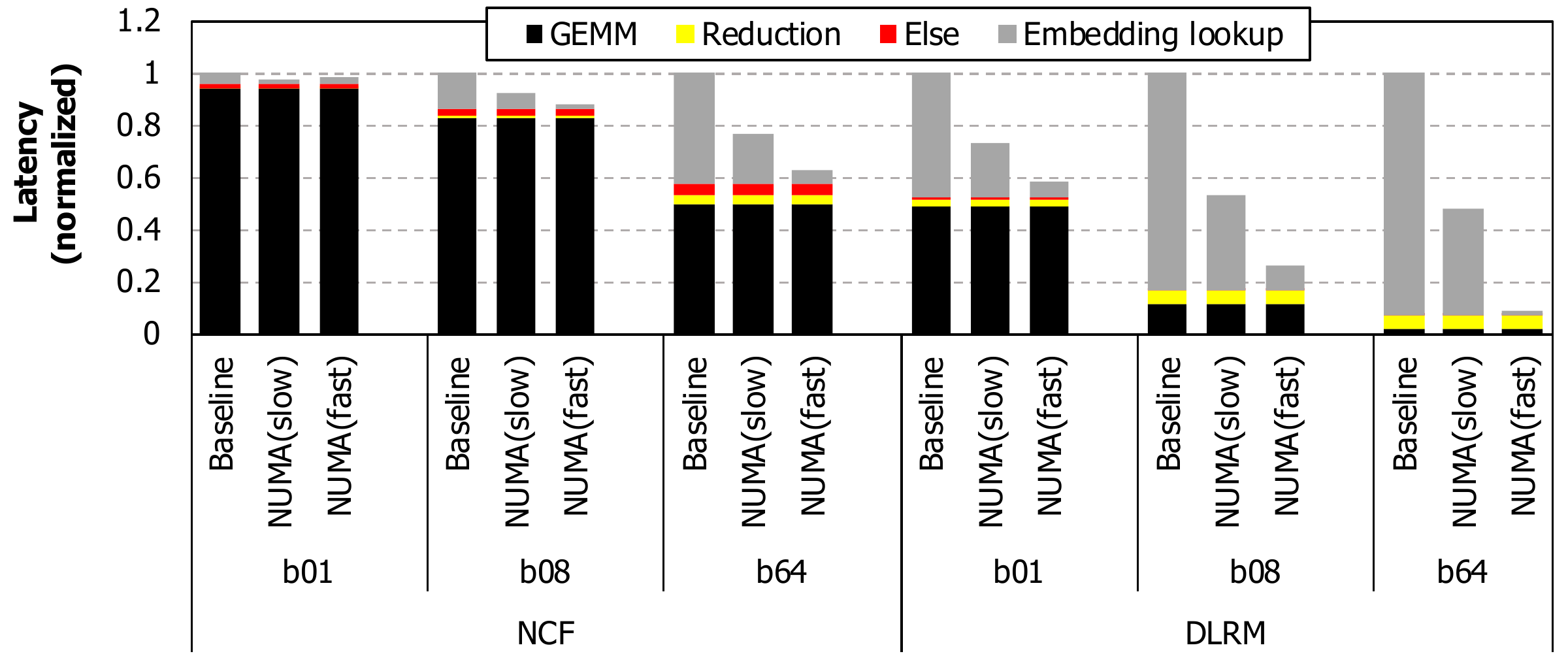}
\caption{
	Latency breakdown when running workloads using embeddings. Baseline MMU-less NPU (\texttt{baseline}) 
		assumes remote embeddings are copied to CPU memory first and then
			copied again to the destination NPU over PCIe.  We study two
		NUMA systems enabled by \nmmu: 1) NUMA over legacy PCIe system interconnect (\texttt{NUMA(slow)}) and 2) NUMA over 
		high-bandwidth GPU/NPU interconnects like NVIDIA's NVLINK~\cite{nvlink} (\texttt{NUMA(fast)}).
We follow prior work on
		NUMA CPU-GPU or multi-GPU studies~\cite{beyond_the_socket,ararwal:asplos:2015,gpu_paging} when modeling the
		added latency and communication bandwidth constraints of CPU$\leftrightarrow$NPU or NPU$\leftrightarrow$NPU interconnects (\tab{tab:npu_config}).
}
\label{fig:numa_vs_baseline}
\end{figure}

As discussed in \sect{sect:embedding}, state-of-the-art 
recommendation systems using embeddings exhibit highly sparse,
							 irregular memory accesses over a large
							 embedding table (\fig{fig:ncf}).  To overcome the memory
							 capacity bottleneck, Facebook's DLRM~\cite{facebook_dlrm} for
							 instance employ an \emph{accelerator-centric} parallelization
							 strategy where the embedding table is model-parallelized across
							 multiple GPUs (in our case the NPUs). This however comes at the
							 cost of an \emph{all-to-all} communication among the GPUs (NPUs)
	to gather all the embeddings from remote memory regions
	(\fig{fig:parallelized_dlrm}).  Because current MMU-less NPUs cannot address
	memory outside its local, physical memory, an intermediate
	solution that facilitates remote embedding gathers will be to have the CPU
	runtime manually copy the (remote) embedding vectors to a buffer allocated in
	CPU memory, which is then copied over to the (local) NPU memory.  Since the
	embedding vectors are sparse with only several hundreds of bytes in size,
	current GPUs can instead utilize its MMU to directly access remote GPU memory
	for embedding gathers in a CC-NUMA fashion or migrate the missing pages into
	its local memory, obviating the need for a CPU-involved, intermediate 	data
	transfer and copy operation.  With our proposed \nmmu architecture in hand,
	the NPU can page-fault on missing pages mapped to a remote NPU's physical
	memory, and either 1) directly fetch the embeddings using fine-grained NUMA
	accesses, or 2) migrate the missing page into its local memory
		(\sect{sect:large_pages}).  \fig{fig:numa_vs_baseline} shows the performance
	advantage of utilizing NUMA accesses for handling sparse embedding layers. As
	depicted, the baseline MMU-less NPU suffers from significant increase in
	latency because of the (redundant) manual data copies over CPU memory. \nmmu
	is able to reduce the latency spent in gathering the embeddings as this
	process is undertaken directly over the (PCIe or NVLINK) system interconnect
	in a fine-grained NUMA fashion, achieving an average $31\%$ and $71\%$
	latency reduction than baseline NPU without an MMU. These results
	highlight the merits of featuring address translations in
	NPUs, which we will be of utmost importance as DL
	workloads evolve into having high capacity demands with irregular
	memory access behaviors. 

\section{Discussion}
\label{sect:discussion}

\subsection{NeuMMU with Large Pages}
\label{sect:large_pages}

\begin{figure}[t!] \centering
\includegraphics[width=0.44\textwidth]{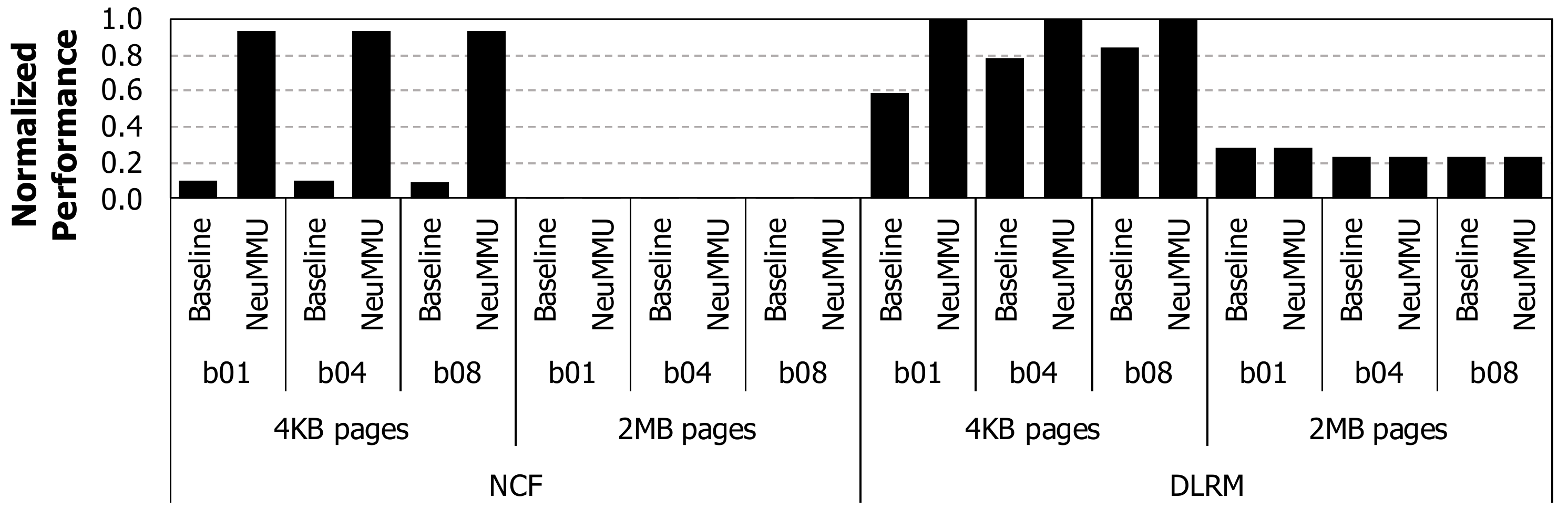}
\caption{
Performance of executing the DL workloads using sparse embedding layers~\cite{he:www:2017,facebook_dlrm} over baseline IOMMU  
and \nmmu, normalized to oracular MMU. The experiment assumes similar evaluation settings as in \fig{fig:numa_vs_baseline} except that 
the missing embeddings are brought into local physical memory using demand paging. 
}
\label{fig:large_page_for_ncf}
\end{figure}

The computation and memory access characteristics of conventional dense
CNNs/RNNs are highly regular with its tensor data (i.e., \ia and
		\weights) sized at several hundreds of MBs. We observe that $2$MB large
pages can substantially decrease the performance overhead of baseline IOMMU
address translations for the dense CNNs/RNNs (average $4\%$, worst case
		$10\%$). Our \nmmu architecture successfully removes such performance
overheads as in the baseline $4$KB pages. Given the high efficiency of IOMMU
based address translation under large pages, one might presume that NPU MMUs
should simply employ large pages exclusively without baseline small
pages.  However, for DL workloads that exhibit sparse data access patterns with
large memory footprint (such as our studied sparse embedding layers), large
pages can incur a much significant performance overhead compared to small
pages.  Large pages increase the data transfer size of each demand paged
request (recall that a single embedding is only hundreds of bytes with low
		temporal/spatial locality, \fig{fig:ncf}) which, not only causes
significant (redundant) communication traffic on the system interconnect and
hurt performance, but also wastes NPU physical memory by causing memory bloats
via internal fragmentation~\cite{kwon:osdi:2016}. Rather than using NUMA to
gather sparse embeddings (as assumed in \sect{sect:numa_npu}),
			 \fig{fig:large_page_for_ncf} summarizes the performance of small and
			 large pages when utilizing \nmmu to \emph{page-fault} and
			 migrate the missing (sparse) data using \emph{demand-paging} into the
			 NPU physical memory. For small pages, \nmmu performs well to recover
			 the lost performance and improves performance from $17\%$ up to an
			 average $96\%$ of oracle. Unfortunately, the performance overhead of
			 large pages cannot be recovered with \nmmu because of the (redundant)
	prefetching effects of large pages over the sparse access patterns of NCF and DLRM.
	These results highlight the importance of providing robust address
	translation service for both small and large pages for current and future DL
	workloads. Prior work by Ausavarungnirun et
	al.~\cite{ausavarungnirun:micro:2017} that synergistically combines small
	and large pages concurrently can be a promising solution to address these
	issues. Nonetheless, our paper focuses on efficient address
	translation support so evaluating efficient page-fault handling and demand
	paging solutions that closes such performance for large pages is beyond our
	scope and we leave it as future work.

\subsection{NeuMMU with Alternative NPU Architectures}
\label{sect:neummu_with_other_npus}

There have been numerous NPU designs proposed in prior literature so it is
challenging to define a ``generic'' NPU architecture that represents all design
points in such fast-evolving space. As such, our baseline NPU
assumed Google's systolic-array microarchitecture as it is to date the most successfully deployed
NPU design. To study the applicability of \nmmu on alternative NPU designs, we 
also developed a cycle-level performance model that follows several representative prior work
based on the \emph{spatial} architecture design
paradigm~\cite{dadiannao,eyeriss,cnvlutin,scnn,park:2018:olaware}. Our spatial-array
based NPU design is modeled similar to DaDianNao~\cite{dadiannao} or Eyeriss~\cite{eyeriss}, which
employs a two-dimensional grid of PEs, each of which contains a vector ALU that
handles dot-product operations. These spatial NPU architectures also employ
an SPM-centric on-chip memory hierarchy, which our \nmmu design is founded upon,
	 and our evaluation showed that our
\nmmu architecture is able to similarly close the performance gap of baseline
IOMMUs, only incurring an average $2\%$ performance overhead. We omit
the results due to space limitations. 

\subsection{Sensitivity}
\label{sect:sensitivity}

The robustness of \nmmu has been studied over several different architecture
configurations as well as different application batch sizes.  For \nmmu design
space exploration, we sweep the number of \prmb mergeable slots ($1$ to $32$),
			parallel PTWs ($64$ to $256$), \tpr entries, and total TLB entries ($128$
					to $2K$). Across all the sensitivity studies, the
			performance achieved was never less than
			$73\%$ with an average $97\%$ of the oracular MMU.  We also studied our
			workloads with large batch size of $32$, $64$, and $128$. As mentioned in
			\sect{sect:eval_methodology}, large batches lead to intractable amount of
			simulation time, so we limit our evaluation to the common layer
			configuration of each DNN. Similar to small batches, the baseline
			IOMMU achieves an average $5.9\%$ of	oracle. \nmmu 
			successfully closes this performance gap for large batches, reaching
			$99.9\%$ of oracular MMU.  These results demonstrate the robustness
			of our throughput-centric \nmmu for SPM-centric NPUs.

\section{Related Work}
\label{sect:related}

Our work builds on top of prior studies on hardware and software mechanisms to accelerate
MMUs. Here we first summarize closely related work on CPU/GPU MMUs, followed by a summary on prior literature
designing ML accelerator architectures.

{\bf Address Translation for CPUs.} As application memory footprint increases,
	commercial CPUs have  started including multi-level TLB
	hierarchies~\cite{multitlb:1,haswell} with per-core MMU caches to accelerate
	the page-table walking process. Barr et al.~\cite{mmu_cache} explored the
	design space of MMU caches, showing that the most effective ones are unified
	translation caches. Prefetching
	translations~\cite{tlbprefetch:1,tlbprefetch:2,tlbprefetch:3}, shared
	TLBs~\cite{shared_tlb}, and MMU caches~\cite{shared_mmu} has also been
	studied in the literature to alleviate translation overheads in various CPU
	context.

{\bf Address Translation for Accelerators.} There have been some pioneering
work by Power et al~\cite{gpu_x86_at} and Pichai et al.~\cite{gpu_tlb} that
explored the benefits of GPUs in utilizing IOMMU for VA-to-PA address
translations.  Both of these studies proposed a per-core, post-coalescer TLB,
dedicated logic to walk the page-tables,
multi-level shared TLB hierarchy, and a page translation cache, similar to our
proposal. Hao et al.~\cite{cong:2017:hpca} studied the utilization of IOMMUs for accelerator-centric
						systems, similarly proposing shared TLBs, parallel PTWs, and MMU caches.
Our work differentiates itself from all these
						prior studies as we specifically target NPUs with a carefully
						designed, throughput-centric MMU that is tailored for DNNs. 
As discussed in \sect{sect:nmmu_proposal}, we quantitatively demonstrated that prior
translation locality-centric MMUs are not able to sufficiently
handle the translation bursts of DNNs. Our study provides the key insight that NPU MMUs
should be designed for enhancing translation throughput, rather than
translation locality, leading to our novel \prmb and \tpr microarchitecture on top of
a massively parallel page-table walker design.

{\bf Accelerator architectures for ML.} Aside from these closely related prior
work on MMUs, there has been a large body of prior work exploring the design of
space of ML accelerator architectures~\cite{diannao,dadiannao,shidiannao,pudiannao,du:2015:micro,minerva,dnn_pim_reram,eyeriss,cambricon,isacc,neurocube,redeye,tabla,dnnweaver,intel:2017:fpl,gao:2017:tetris,intel:2018:fpga,rhu:2016:vdnn,mcdla:cal,mcdla,tensordimm,choi:2020:prema,kwon:2019:disagg}
with recent interest on sparsity-optimized solutions for further
energy-efficiency
improvements~\cite{song:2015:eie,cnvlutin,cambriconx,stripes,bitpragmatic,intel:2017:icassp,intel:2017:fpga,whatmough:2017:isscc,whatmough:2017:hotchips,scnn,bittactical,rhu:2018:cdma}.
Our work on NPU MMUs is orthogonal to these prior art as our primary focus is on adding new \emph{features} to these ML accelerator designs. 

\section {Conclusion}
\label{sect:conclusion}

As the computation demands for DL workloads increase, we expect NPUs to evolve
into first-class citizens in heterogeneous computing platforms. We make a case
for providing address translation capabilities for NPUs, an important first
step in evolving these devices as primary compute engines. Through a
data-driven application characterization study, we root-cause the challenges in
prior GPU-centric MMU solutions and propose three novel architecture designs
tailored for the application behavior of DNNs. Compared to an oracular MMU
design, our proposal achieves only an average $0.06\%$ performance overhead
while allowing CPUs and NPUs to share a unified global address space.


\bibliographystyle{ieeetr}
\bibliography{ref}

\end{document}